\documentclass[twocolumn,pra,superscriptaddress]{revtex4-1}
\usepackage{graphicx,graphics,bbm,amsfonts,amsmath,amsthm,mathrsfs,amssymb}
\usepackage{hyperref}
\usepackage{multirow}
\usepackage{slashed}

\newcommand{\upe}{\mathrm{e}}
\newcommand{\upi}{\mathrm{i}}
\newcommand{\upd}{\mathrm{d}}
\newcommand{\Tr}{\mathrm{Tr}}
\newcommand{\rhoS}{\rho_\mathrm{S}}
\newcommand{\rhoB}{\rho_\mathrm{B}}
\newcommand{\id}{\mathbbm{1}}
\newcommand\cE{\mathcal{E}}
\newcommand\cJ{\mathcal{J}}
\newcommand\cN{\mathcal{N}}
\newcommand\cT{\mathcal{T}}
\newcommand\cU{\mathcal{U}}
\newcommand\sS{\mathscr{S}}
\newcommand{\HStar}{H_\mathrm{tar}}
\newcommand{\HSsim}{H_\mathrm{sim}}
\newcommand{\Htar}{H^\mathrm{(tar)}}
\newcommand{\Hsim}{H^\mathrm{(sim)}}
\newcommand{\HS}{H_\mathrm{S}}
\newcommand{\HB}{H_\mathrm{B}}
\newcommand{\HSB}{H_\mathrm{SB}}
\newcommand{\UStar}{U_\mathrm{tar}}
\newcommand{\USsim}{U_\mathrm{sim}}

\newcommand{\Tar}{\mathrm{tar}}
\newcommand{\Sim}{\mathrm{sim}}
\newcommand{\Uerr}[1]{\cU_{\mathrm{err},#1}}
\newcommand{\Err}[1]{\mathrm{Err}_{#1}}
\newcommand{\tauB}{\tau_\mathrm{B}}
\newcommand{\epmax}{\epsilon_{\max}}
\newcommand{\ommax}{\omega_{\max}}

\begin{document}
\title{A digital quantum simulator in the presence of a bath}
\author{Yi-Cong Zheng}
\email{zheng.yicong@quantumlah.org}
\affiliation{
Centre for Quantum Technologies, National University of Singapore, Singapore}
\affiliation{
Yale-NUS College, Singapore}
\author{Hui-Khoon Ng}
\email{huikhoon.ng@yale-nus.edu.sg}
\affiliation{
Centre for Quantum Technologies, National University of Singapore, Singapore}
\affiliation{
Yale-NUS College, Singapore}
\affiliation{
MajuLab, CNRS-UNS-NUS-NTU International Joint Research Unit, UMI 3654,
Singapore}
\date{\today}

\begin{abstract}
For a digital quantum simulator (DQS) imitating a target system, we ask the following question: Under what conditions is the simulator dynamics similar to that of the target in the presence of coupling to a bath? In this paper, we derive conditions for close simulation for three different physical regimes, replacing previous heuristic arguments on the subject with rigorous statements. In fact, we find that the conventional wisdom that the simulation cycle time should always be short for good simulation need not always hold up. Numerical simulations of two specific examples strengthen the evidence for our analysis, and go beyond to explore broader regimes.
\end{abstract}
\maketitle

\section{Introduction}
Quantum simulators have attracted a lot of interest, both theoretical and experimental \cite{Feynman:1982:467,Lloyd:1996:1073,nori_simulator_science,nori2014rev_quantumsimulator}. Theoretical understanding of the potential of quantum simulators in addressing problems beyond the reach of classical computations remains incomplete, but quantum simulators present much nearer-term experimental goals than full-fledged quantum computers.
There are two general classes of quantum simulators: \emph{analog}~\cite{kendon2010quantum_analog_computing} and \emph{digital}~\cite{Lloyd:1996:1073}. Analog simulators are devices whose Hamiltonians can be engineered to imitate a target model continuously in time;
digital quantum simulators (DQS), on the other hand, stroboscopically approximates the time evolution of the target system by applying a discrete sequence of gates. The latter is the subject of this article.

Different physical systems and architectures have been considered as platforms for quantum simulation, with different targets in mind. Theoretical proposals using Rydberg atoms \cite{RydbergsimulatorNatPhysics,weimer2011digital_rydberg_simulator} and experimental demonstrations with trapped ions \cite{ion_trap_open_2011_nature,muller2011simulating_ion,ion_trap_gauge_field_2016_nature} have explored the simulation of both closed quantum systems as well as open systems with Markovian dynamics. The possibility of simulating non-Markovian dynamics with DQS was suggested in Ref.~\cite{digital_simulator_non_markovian}. Another desirable target is to simulate many-body Hamiltonians that provide natural tolerance to noise. An example is the four-body Kitaev toric code model \cite{Kitaev:2003:2}, with a degenerate ground space in which stored information is protected from leakage into the excitation space by an energy gap large compared to the energy scale of the noise. Such models provide the foundations for schemes for quantum memory~\cite{Dennis:2002:4452,brown2016quantum_memory,terhal2015quantum}, adiabatic quantum computation~\cite{FarhiScience,JordanFarhiShorPhysRevA.74.052322}, fault-tolerant quantum computation~\cite{Nielsen:2000:CambridgeUniversityPress,Yi-Cong_PhysRevA.89.032317,zheng2015fault_hqc_surface, cesare2015adiabatic}, and topological quantum computation~\cite{Kitaev:2003:2, Nayak:2008:1083}. 

Our work focuses on this last goal of simulating many-body Hamiltonians for natural noise tolerance. Because of the many-body nature, the desired Hamiltonians are usually difficult to realise exactly in the lab. Instead, digital simulation is used to achieve an \emph{effective} Hamiltonian resembling the target. Keeping in mind that the target Hamiltonian is chosen for its tolerance to the noise from the enviroment, the criterion for close and useful simulation between the DQS and the target has to include a comparison of not just the system-only Hamiltonian, but also the noise seen by the simulator and the target. One might imagine that a simulator can achieve close simulation of the target Hamiltonian, but the noise as seen by the simulator, due to the gate sequences, becomes different from the one against which the target provides natural resilience. A close simulation of the target Hamiltonian of this sort can hardly be considered to have achieved its original goal of protection against noise. 

We thus address the question: Under what physical conditions do we have close simulation of the target Hamiltonian \emph{and} dynamics in the presence of the bath, which is the source of noise?
Conventional wisdom \cite{lloyd1999robust,young2012finitetemperature_toric,becker2013dynamic_self_correction} tells us that, heuristically, fast gates and short simulation cycles should suffice. Here, we do a careful analysis, and derive the precise conditions for close simulation. It turns out that the short simulation cycle alone is neither necessary nor sufficient. Surprisingly, one can find circumstances that demand a longer cycle for better simulation.

The structure of this paper is as follows. In Sec.~\ref{sec:TarVsSim}, we introduce basic concepts and formulate the problem. In Sec.~\ref{sec:analytics}, we quantify the simulation error analytically under different physical regimes, and derive the conditions for good simulation. Section~\ref{sec:numerics} numerically addresses specific examples---that of a toric-code vertex, and of a five-qubit code---to examine regimes inaccessible to the analysis of Sec.~\ref{sec:analytics}. We close with a summary and discussion of our results in Sec.~\ref{sec:Conc}. To help the reader with the notation used throughout the article, Appendix \ref{app:Glossary} gathers a glossary of the symbols used.

\clearpage

\section{Target versus simulator dynamics}\label{sec:TarVsSim}

Consider a controllable system $\mathrm{S}$ evolving jointly with a bath $\mathrm{B}$---the source of noise for $\mathrm{S}$---according to the Hamiltonian
\begin{equation}
H=\HS+\HB+\alpha\HSB.
\end{equation}
$\HS$ is the natural (in contrast with the modified versions below) system-only Hamiltonian, $\HB$ is the bath-only Hamiltonian, and $\HSB$ is the system-bath interaction, accompanied by a book-keeping parameter $\alpha$. The system-bath interaction is assumed to be weak, a precondition for $\mathrm{S}$ to be useful for quantum information processing tasks, enforced by regarding $\alpha\ll1$.

\subsection{Simulating a target system Hamiltonian}

\begin{figure}[!ht]
\centering\includegraphics[width=\columnwidth]{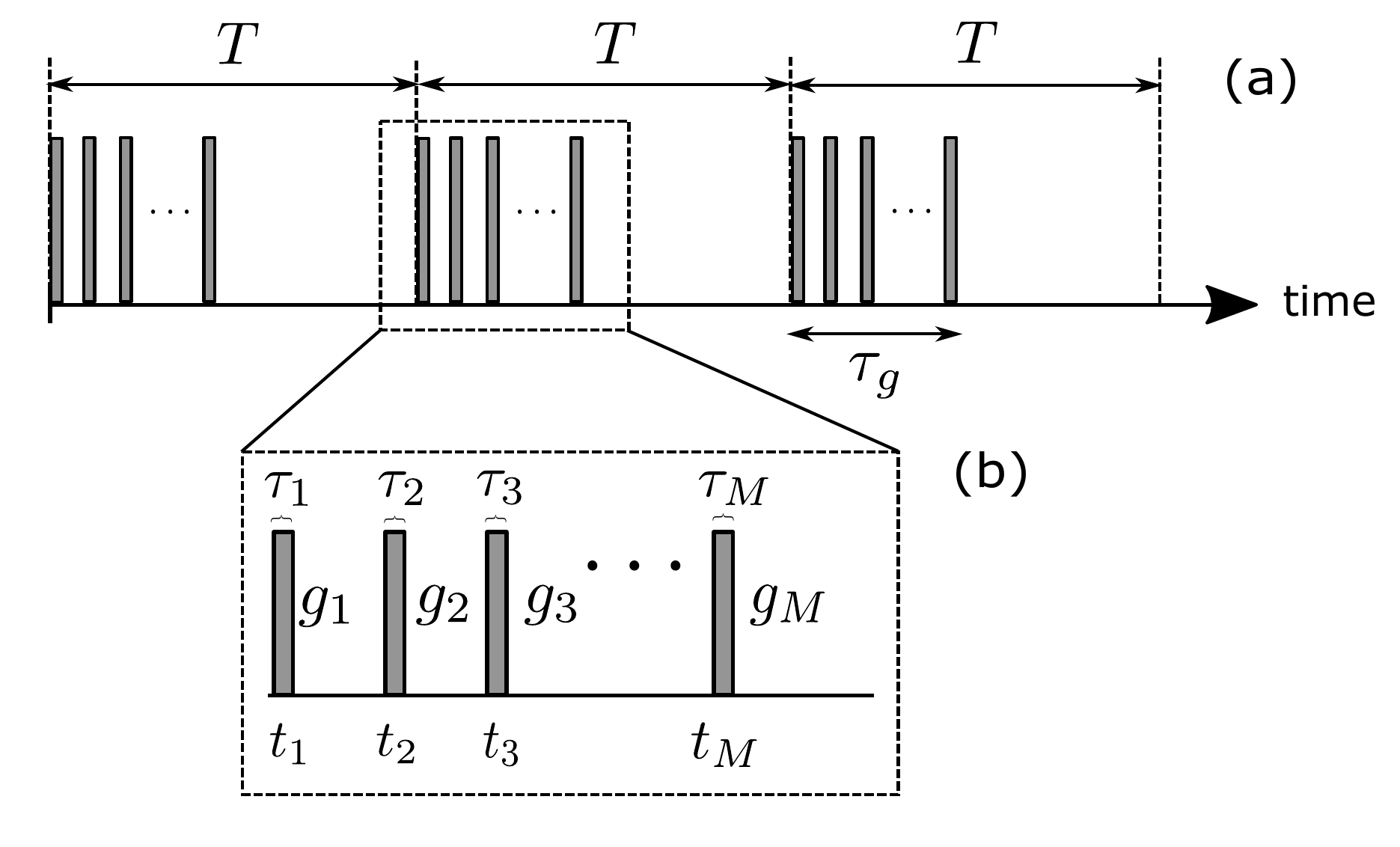}
\caption{\label{fig:gate_sequence} A periodic pulse sequence, with cycle time $T$. Each pulse sequence comprises a set of $M$ pulses that altogether take time $\tau_g$ to complete. The $i$th pulse, implementing gate $g_i\equiv \upe^{-\upi H_{g_i}\tau_i}$, starts at time $t_i$, and lasts for duration $\tau_i$.
}
\end{figure}

For the moment, let us forget about the bath, and focus on the system.
The idea of a quantum simulator is to modify the natural dynamics of the system to one that follows a target system Hamiltonian $\HStar$, which we assume to be time-independent. In a DQS, one achieves \emph{stroboscopic} simulation by applying a periodic sequence of short pulses on $\mathrm{S}$, each pulse implementing a particular unitary gate operation. Assuming that the natural system Hamiltonian $\HS$ is time-independent, the DQS evolves according to a piecewise-constant (system-only) simulator Hamiltonian,
\begin{equation}
\HSsim(t) = \sum_i \,\Bigl[\Theta(t-t_i)-\Theta\bigl(t-(t_i+\tau_i)\bigr)\Bigr] \,H_{g_i}\,.
\end{equation}
Here, $\Theta(\,\cdot\,)$ is the Heaviside step function, $t_i$ is the starting time of the $i$th pulse with strength $H_{g_i}\!-\HS$, and $\tau_i$ is the duration of the pulse, taken to be $\tau_p$ for all $i$ for simplicity. A sequence with $M$ gates is  illustrated in Fig.~\ref{fig:gate_sequence}.

Each cycle of the periodic pulse sequence takes total time $T$, so $\HSsim(t+T)=\HSsim(t)$. That the DQS simulates the target is encapsulated by the \emph{simulation condition},
\begin{equation}
\USsim\!\bigl((N\!+\!1)T,NT\bigr)\simeq \UStar\!\bigl((N\!+\!1)T,NT\bigr)~\forall N\!\!\in\!\mathbb{Z}_0^+.
\end{equation}
Here, $U_\mu(t',t)$, for $\mu\equiv \Tar$  or $\Sim$, is the unitary evolution operator,
\begin{equation}\label{eq:Umu}
U_\mu(t',t)\equiv  \cT_+\exp{\left(-\upi\int_t^{t'}\upd s\, H_{\mu}(s) \right)}\,,
\end{equation}
for the target or the simulator. We use units where $\hbar=1$.
If $\USsim\bigl((N+1)T,NT\bigr)= \UStar\bigl((N+1)T,NT\bigr)$ for all $N\in\mathbb{Z}_0^+$, we say that the simulator is \emph{exact}. Note that the periodicity of $\HSsim$ means that the simulator is exact if and only if the simulation condition is satisfied with an equality for $N=0$.
More typically, the simulator is not exact and there is a nonvanishing \emph{design error} (see below for a precise definition).

A good DQS behaves like the target stroboscopically, at the completion of every cycle of the pulse sequence, but there is no requirement for close simulation at other times. For close simulation, the cycle time $T$ should be short compared to the timescales of the target system, so that the features of the target are faithfully reproduced in the simulator \footnote{Afterall, one would hardly say that the constant zero function is close to the sine function even though they have the same value every half-cycle of the sine. Instead, one aims for an approximation with similar values to the sine at intervals small compared to its period.}. The timescales of the target are determined by the set of transition frequencies $\{\omega:\omega= \varepsilon-\varepsilon'\}$, where $\varepsilon$ and $\varepsilon'$ are eigenfrequencies of $\HStar$. Close simulation hence requires, and we will assume this throughout the article,
\begin{equation}\label{cond:shortT}
\ommax\, T\ll 1,
\end{equation}
where $\ommax\equiv\max|\omega|$, the largest (absolute value of the) transition frequency of the target system.

Condition \eqref{cond:shortT} anyway underlies the Trotter-Suzuki-type decomposition often used in the simulator gate-sequence design. Consider a target Hamiltonian of the form
$\HStar=\sum_{\ell=1}^L h_\ell$, a sum of generally noncommutting terms. A concrete example would be a square lattice with qubits located on the edges, and $h_\ell$ are local four-body vertex or plaquette operators; the toric code would have commuting $h_\ell$s, but one could imagine other examples.
A simple design of $\HSsim$ is to employ a Trotter-Suzuki decomposition to approximate $\UStar(T,0)=\upe^{-\upi\HStar T}$:
\begin{equation}\label{eq:Trotter}
\upe^{-\upi\HStar T} \simeq \upe^{-\upi h_LT}\cdots\,\upe^{-\upi h_2T}\,\upe^{-\upi h_1T} + O{\Bigl({\bigl(\sum_\ell\Vert h_\ell\Vert T\bigr)}^2\Bigr)}.
\end{equation}
The error $O(\cdot)$ can be pushed to higher order with more complicated decompositions~\cite{wiebe2011simulating}, but all demand satisfaction of Condition \eqref{cond:shortT} for good approximation.

\subsection{Dynamics in the presence of the bath}
\begin{figure}[!ht]
\centering\includegraphics[width=0.6\columnwidth]{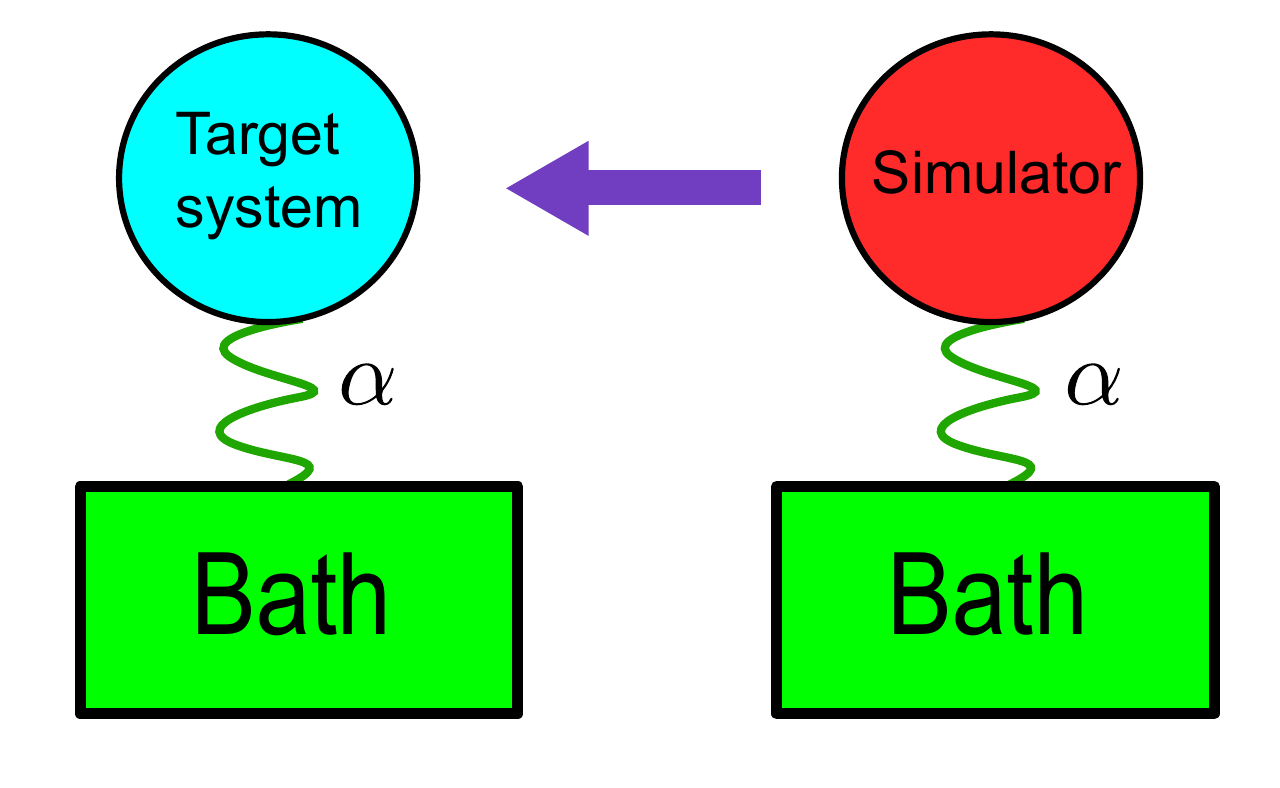}
\caption{\label{fig:illustrator} The simulator and the target system interact with the bath in the same manner.}
\end{figure}

The goal in many quantum simulation scenarios is to have $\HStar$ provide passive resilience against the noise due to the unavoidable bath coupling~\cite{lloyd1999robust,JordanFarhiShorPhysRevA.74.052322,young2012finitetemperature_toric}. The target $\HStar$ is usually designed for the natural noise seen by $\mathrm{S}$ for the given $\HSB$, e.g., by choosing an $\HStar$ with a ground-state manifold protected by an energy gap large compared to the energy scale set by the $\HSB$ coupling. For this to work, the implicit assumption is that the simulator, upon interaction with the bath, behaves similarly to the open target system, i.e.,
that the dynamics of the simulator under the joint Hamiltonian (see Fig.~\ref{fig:illustrator})
\begin{equation}
\Hsim=\HSsim(t) +\HB+\alpha\HSB
\end{equation}
resemble that of the target system under
\begin{equation}
\Htar=\HStar+\HB+\alpha\HSB.
\end{equation}

While $\HB$ and $\HSB$ are themselves unchanged---in fact, unchangeable---by the simulator pulse sequences, the \emph{dynamics} of the system depend on the interplay between $\HSsim$, $\HB$ and $\HSB$, and their relative timescales. There is hence no \emph{a priori} reason to expect a close simulation of $\HStar$ by $\HSsim$ to guarantee a close simulation of the system dynamics in the presence of the bath. Because of the stroboscopic nature of the DQS, one expects any noise process that occurs on a timescale faster than the period $T$ to notice a difference between $\HStar$ and $\HSsim$, but that need not be the only condition for close simulation; the details are rather more intricate, as we shall see.

\subsection{The simulation error}
We are interested only in the dynamics of the target and simulator systems, not that of the bath. The relevant quantity is then the quantum channel that takes the system from the initial time $t=0$ to some time $t>0$, i.e., the joint evolution according to $H^{(\mu)}$, for time $t$ from the initial joint system-bath state, followed by a partial trace on the bath. The stroboscopic nature of the simulation suggests a comparison of target and simulator at times $t=NT$ with $N$ a nonnegative integer. Specifically, we look at the (completely positive, trace-preserving) channel on the system only after time $t=NT$,
\begin{equation}
\cE^{(\mu)}_{N}(\,\cdot\,)\equiv\Tr_B{\left\{U^{(\mu)}(NT,0)\left(\,\cdot\,\otimes\rho_B\right) U^{(\mu)}(NT,0)^\dag\right\}},
\end{equation}
for $\mu\equiv\mathrm{tar},\mathrm{sim}$. Here, $U^{(\mu)}(t',t)$ is the unitary evolution operator for $H^{(\mu)}$,
\begin{equation}
U^{(\mu)}(t',t)\equiv  \cT_+\exp{\left(-\upi\int_t^{t'}\upd s\, H^{(\mu)}(s) \right)},
\end{equation}
for the joint target-bath or simulator-bath time evolution. $\cT_+$ is the time-ordering operator. For the various unitary evolution operators, we will use the shorthand of $U(t)\equiv U(t,0)$ for evolution from the initial time $t=0$. Here, we have taken the initial system and bath state to be a product state, a good approximation in typical quantum information processing situations.

We define the \emph{simulation error} after $N$ cycles to be the difference between the target and simulator channels,
\begin{equation}
\Err{N}\equiv {\left\Vert \cE^{(\mathrm{tar})}_N-\cE^{(\mathrm{sim})}_N\right\Vert}.
\end{equation}
Here, $\Vert\cdot\Vert$ is a unitarily-invariant norm.
One can better understand this simulation error by going into the interaction picture defined by $H^{(\mu)}_0\equiv H_\mu+\HB$, with the associated unitary evolution operator
\begin{equation}
U_0^{(\mu)}(t',t)=U_\mu(t',t)\otimes U_B(t',t),
\end{equation}
where $U_\mu(t',t)$ is the system-only operator as defined in Eq.~\eqref{eq:Umu} and $U_B(t',t)\equiv \upe^{-\upi H_B(t'-t)}$, assuming a time-independent $H_B$. One can then write $U^{(\mu)}$ as
\begin{equation}
U^{(\mu)}(t',t)=U^{(\mu)}_0(t',t)U^{(\mu)}_I(t',t),
\end{equation}
with $U_I^{(\mu)}$ the interaction-picture evolution operator,
\begin{equation}
U_I^{(\mu)}(t',t) = \cT_+\exp{\left(-\upi\int_t^{t'} \upd s\, \HSB^{(\mu)}(s) \right)},
\end{equation}
for $\HSB^{(\mu)}(s) \equiv U_0^{(\mu)}(s)^\dagger \HSB U_0^{(\mu)}(s)$. Then, $\cE_N^{(\mu)}$ can be rewritten as
\begin{eqnarray}
&&\cE^{(\mu)}_{N}(\,\cdot\,)\\
&=&\Tr_B{\left\{U_\mu(nT)U_I^{(\mu)}(NT)\left(\,\cdot\,\otimes\rhoB\right) U_I^{(\mu)}(NT)^\dag U_\mu(NT)^\dagger\right\}}.\nonumber
\end{eqnarray}
Under the unitarily-invariant norm, the simulation error is
\begin{eqnarray}
\Err{N}={\left\Vert\mathcal{E}^{(\Tar)}_{N,I}-\Uerr{N}\circ\mathcal{E}^{(\Sim)}_{N,I}\right\Vert}
\end{eqnarray}
where
\begin{equation}\label{eq:EI}
\cE^{(\mu)}_{N,I}(\,\cdot\,)\equiv\Tr_\mathrm{B}{\left\{U^{(\mu)}_I(NT)(\,\cdot\,\otimes \rhoB) U^{(\mu)}_I(NT)^\dagger\right\}},
\end{equation}
and
\begin{equation}
\Uerr{N}(\cdot)
\!\equiv\!{\left[\UStar\!(NT)^\dagger \USsim\!(NT)\right]} (\cdot){\left[\UStar\!(NT)^\dagger \USsim\!(NT)\right]}^\dagger
\end{equation}
captures the \emph{design error} after $N$ cycles: Its deviation from the identity channel is due solely to the chosen pulse sequence. An exact simulator has no design error, i.e., $\Uerr{N}=1$ the identity map, and its simulation error is simply the difference between $\cE^{(\mathrm{tar})}_{N,I}$ and $\cE^{(\mathrm{sim})}_{N,I}$, arising only from the system-bath coupling.

\subsection{Notation}
Before we proceed further, we collect here a few general remarks to help the reader with the notation used throughout the text. For a real number $a$, $\lfloor a\rfloor$ denotes the ``floor'' of $a$, i.e., the largest integer less than or equal to $a$. A slashed symbol refers to the fractional part of the quantity, e.g., $\slashed{a}\equiv a-\lfloor a \rfloor \in[0,1)$. For a real number $y$, $[y]_+\equiv y\,\Theta(y)$, where $\Theta(y)$ is the step function, i.e., $\Theta(y)$ is $1$ if $y\geq 0$, and is $0$ if $y<0$.

We measure time in units of the stroboscopic period $T$, and frequencies in units of $1/T$. A tilde atop a function refers to the dimensionless version of that function (as defined in the text), e.g., $\widetilde f_{k\ell}(\cdot)$ is the dimensionless version of $f_{k\ell}(\cdot)$. The letters $s$ and $t$ (and their primed versions) are time quantities; the letters $a$, $b$ and $c$ are the dimensionless (i.e., measured in units of $T$) counterparts, e.g., $a=s/T$. $\omega$ and $\nu$ are frequencies; $\epsilon\equiv \omega T$ and $x\equiv \nu T$ are their dimensionless versions.

Quantities with a superscript $(\mu)$ [e.g., $H^{(\Sim)}$ or $U^{(\Tar)}(t)$] contain contributions from the system-bath coupling $\HSB$; those with a subscript $\mu$ [e.g., $H_\Sim$ or $U_\Tar(t)$] contain only the system Hamiltonian $\HStar$ or $\HSsim$, and $\HSB$ does not enter.

A glossary is provided in Appendix~\ref{app:Glossary} to help the reader with the various symbols used in the text.

\section{Analytical estimates}\label{sec:analytics}

Since we are concerned with the simulation error due to the presence of the system-bath coupling, for simplicity, we assume an exact simulator, so that the design error plays no role. In practice, any simulation scheme will have some nonzero design error, but such an error can be reduced by better---if more elaborate---choice of simulation pulse sequences. We thus focus only on the difference $\Err{N}=\Vert \cE^{(\mathrm{tar})}_{N,I}-\cE^{(\mathrm{sim})}_{N,I}\Vert$ that arises from the unavoidable system-bath coupling.

The weak system-bath coupling justifies an analysis perturbative in $\alpha$. We expand $U^{(\mu)}_I(t)$ to second order in $\alpha$:
\begin{eqnarray}\label{eq:UI}
U_I^{(\mu)}(t)&\simeq& \id -\upi\alpha\int_0^t \upd s \HSB^{(\mu)}(s)\\
&&\quad -\alpha^2\int_0^t \upd s\int_0^s \upd s' \HSB^{(\mu)}(s)\HSB^{(\mu)}(s').\nonumber
\end{eqnarray}
Let us write $\HSB=\sum_kA_k\otimes B_k$, where $A_k$ acts on the system, and $B_k$ on the bath, both Hermitian operators. We regard $A_k$s as dimensionless operators, while $B_k$s carry the dimension of frequency (setting $\hbar=1$). Both $A_k$ and $B_k$ are taken to be operators with norm of order 1 so that the strength of the system-bath interaction is captured by the $\alpha$ parameter alone.
Define $A_k^{(\mu)}(t)\equiv U_\mu(t)^\dagger A_k U_\mu(t)$, and $B_k(t)\equiv U_\mathrm{B}(t)^\dagger B_k U_\mathrm{B}(t)$, the interaction-picture operators, so that $\HSB^{(\mu)}(t)=\sum_kA_k^{(\mu)}(t)\otimes B_k(t)$.
Let $\langle B \rangle\equiv \Tr\{B\rhoB\}$, for any bath-only operator $B$, and $\rhoB$ is the initial bath state. We denote the two-point bath correlation functions as
\begin{equation}
f_{k\ell}(t,s)\equiv\langle B_k(t)\,B_\ell(s)\rangle,
\end{equation}
with dimensions of (frequency)$^2$. 
Observe that $f_{k\ell}(t,s)^*=f_{\ell k}(s,t)$.
We make the often-applicable assumption that $\rhoB$ is a stationary state of $\HB$, i.e., $[\HB,\rhoB]=0$, and that $\langle B_k(t)\rangle=0$ $\forall k,t$. Stationarity means that $f_{k\ell}(t,s)=f_{k\ell}(t-s,0)\equiv f_{k\ell}(t-s)$ and $f_{k\ell}(t)^*=f_{\ell k}(-t)$.

In Appendix \ref{app:EDiff}, we show that the difference $\cE^{(\Tar)}_{N,I}-\cE^{(\Sim)}_{N,I}$, to lowest-order in $\alpha$, is a sum of three maps,
\begin{equation}\label{eq:channel_difference}
\cE^{(\Tar)}_{N,I}-\cE^{(\Sim)}_{N,I}=\alpha^2(\Delta_1+\Delta_2+\Delta_3),
\end{equation}
where
\begin{eqnarray}
\Delta_1(\,\cdot\,)&=&\sum_{k\ell}\int_0^{N}\upd b\,{\left\{\Bigl[\Lambda_{k\ell}(b)+\overline{\Lambda}_{k\ell}(b)\Bigr](\,\cdot\,)\widetilde A_k^{(\Tar)}(b)\right.}\nonumber\\
&&~~\qquad\qquad{\left.+\widetilde A_k^{(\Sim)}(b)\,(\,\cdot\,)\Bigl[\Lambda_{k\ell}(b)+\overline{\Lambda}_{k\ell}(b)\Bigr]^\dagger\right\}},\quad\nonumber\\
\Delta_2(\,\cdot\,)&=&-\!\sum_{k\ell}\!\int_0^{N}\!\!\!\!\!\upd b{\left[\widetilde A_k^{(\Tar)}\!(b)\Lambda_{k\ell}(b)\!+\!\overline\Lambda_{k\ell}(b)^\dagger\widetilde A_k^{(\Sim)}\!(b)\right]}(\cdot),\nonumber\\
\label{eq:Delta12_original}
\Delta_3(\,\cdot\,)&=&{\left[\Delta_2(\,\cdot\,)\right]}^\dagger.
\end{eqnarray}
Here, we have switched to dimensionless quantities for a cleaner analysis:
$\widetilde A_{k}^{(\mu)}(a)\equiv A_{k}^{(\mu)}(aT)$, and
\begin{eqnarray}\label{eq:Lambdas}
\Lambda_{k\ell}(b)&\equiv& \int_0^{b}\upd a\,\widetilde f_{k\ell}(b-a)\,{\left[\widetilde A_\ell^{(\Tar)}(a)-\widetilde A_\ell^{(\Sim)}(a)\right]}\nonumber\\
\overline\Lambda_{k\ell}(b)&\equiv& \int_b^{N}\upd a\,\widetilde f_{k\ell}(b-a)\,{\left[\widetilde A_\ell^{(\Tar)}(a)-\widetilde A_\ell^{(\Sim)}(a)\right]},\quad
\end{eqnarray}
with $\widetilde f_{k\ell}(a)\equiv T^2 f_{k\ell}(s\equiv aT)$, the dimensionless correlation function. The integration variables $a$ and $b$ are to be thought of as dimensionless time quantities. Observe that $\Lambda_{k\ell}$ and $\overline\Lambda_{k\ell}$ differ only in their integration limits. 

For our analysis below, it is useful to express the correlation function $\widetilde f$ in terms of its Fourier transform $\widetilde J$, which we refer to as the spectral function,
\begin{equation}
\widetilde f_{k\ell}(a)\equiv \int_{-\infty}^\infty\!\!\upd x \,\widetilde J_{k\ell}(x)\,\upe^{-\upi xa}.
\end{equation}
$\widetilde J_{k\ell}(x)$ is assumed to be significant for $x$ around some central value $\bar x$ (not necessarily the mean), within a width $x_c$ ($\geq\!\!0$), i.e., $\widetilde J_{k\ell}(x)$ is negligible for $|x-\bar x|\gtrsim x_c$, for any $k,\ell$. In terms of the original dimensional quantities, $\bar x=\bar\nu T$ for the central frequency $\bar\nu$, and $x_c=\nu_c T$ for the cutoff frequency $\nu_c$ of $J_{k\ell}(\nu)$, the dimensional spectral function, defined by $f_{k\ell}(s)\equiv\int_{-\infty}^\infty \upd \nu J_{k\ell}(\nu)\upe^{-\upi\nu s}$. $\widetilde J_{k\ell}(x)$ and $J_{k\ell}(\nu)$ are related as $\widetilde J_{k\ell}(x\equiv\nu T)=TJ_{k\ell}(\nu)$. A prototypical example is a spectral function of the form
\begin{equation}\label{eq:Jtilde}
\widetilde J(x)\propto (x-\bar x)^w\upe^{-|x-\bar x|/x_c}.
\end{equation}
In many physical situations, $\bar x=0$, so that one has a (dimensionless) frequency distribution that increases from $x=0$ till around $|x|=x_c$, and thereafter an exponential decay sets in. $x_c$ characterizes the width of $\widetilde J(x)$, or equivalently, $\nu_c\equiv x_c/T$ measures the frequency-width of the dimensional $J(\nu)$. Its inverse gives the time-width of $f(s)$, often referred to as the bath correlation time $\tauB\equiv 1/\nu_c$.

If the target and simulator are identical \emph{at all times}, not just at stroboscopic times $t=NT$, we would have $\widetilde A^{(\Tar)}_\ell(a)=\widetilde A^{(\Sim)}_\ell(a)$ for all $a$, and $\Lambda_{k\ell}$ and $\overline{\Lambda}_{k\ell}$ would vanish, as would the difference $\cE^{(\Tar)}_{N,I}-\cE^{(\Sim)}_{N,I}$. The crux hence lies in bounding the difference between $\Lambda_{k\ell}$ and $\overline{\Lambda}_{k\ell}$ when $t\neq NT$.

The perturbative treatment yields $\Err{N}\sim\alpha^2$ for an exact simulator, which is small if the system-bath coupling is weak, as is necessary for a useful physical implementation of a simulator. A stronger simulation criterion, however, is desirable: that a simulator with a shorter stroboscopic cycle time $T$ compared to other timescales of the problem should have a smaller $\Err{N}$. Since $T$ is a controllable parameter in the simulator, this presents the possibility of tuning the open-system simulation error to be as small as desired, independent of the size of $\alpha$. In the following subsections, we examine the conditions under which this behavior holds. Specifically, we look for situations that guarantee that $\tfrac{1}{\alpha^2}\Err{N}$ is small.

\subsection{Single-gate exact simulator}\label{sec:1gateDQS}
We first consider a simple exact simulator $\sS_1$, with
\begin{equation}
\HSsim(t)=\left\{\begin{array}{ll}
\HStar\frac{T}{\tau_g}&,~t\in[NT,NT+\tau_g]\\
0&,~t\in\bigl(NT+\tau_g,(N+1)T\bigr)
\end{array}\right.,
\end{equation}
for $N\in\mathbb{Z}_0^+$.
$\sS_1$ has one ($M=1$) gate pulse per cycle time $T$, of strength $\HStar T/\tau_g$, that lasts for time $\tau_g\leq T$. Its unitary evolution operator is such that $\USsim(NT)=\UStar(NT)$ for all $N\in\mathbb{Z}_0^+$. $\sS_1$ is exact as $\HSsim$ is simply $\HStar$ with a larger strength so that it need only be applied for a shorter time; but $\USsim(t)\neq\UStar(t)$ for all $ t\neq NT$. Such a simulator, though unrealistic---if one could apply $\HStar$ directly, there is no need for the simulator---allows us to zoom in on the effects of the stroboscopic nature of the simulation, without having to worry about the precise pulse sequence used.

For time $s$, the unitary evolution operator for $\mathscr{S}_1$ is
\begin{equation}
\USsim(s)=\upe^{-\upi\HStar{\left(\lfloor s/T\rfloor +1\right)}T}\upe^{\upi\HStar T{\left[1- \frac{1}{\tau_g}(s-\lfloor s/T\rfloor T)\right]}_+}.
\end{equation}
We write $\HStar$ in its eigendecomposition: $\HStar = \sum_\varepsilon \varepsilon P_\varepsilon$, where $\varepsilon$s are the eigenvalues of $\HStar$, and $P_\varepsilon$s are the projectors onto the $\varepsilon$-eigenspaces. Then, the interaction-picture $A$-operators are
\begin{equation}
A_\ell^{(\mu)}(s)=\sum_\omega \upe^{-\upi\omega T(\lfloor s/T \rfloor+1)}\upe^{\upi\omega T{\left[1-\frac{1}{\tau_\mu}(s-\lfloor s/T\rfloor T)\right]}_+}A_\ell(\omega),
\end{equation}
with $A_k(\omega)\equiv\sum_{\varepsilon^\prime-\varepsilon=\omega}P_\varepsilon A_k P_{\varepsilon^\prime}$, $\tau_{\mathrm{tar}}\equiv T$, and $\tau_{\mathrm{sim}}\equiv \tau_g$.
Switching to dimensionless quantities, we have
\begin{eqnarray}
\widetilde A_\ell^{(\mu)}(a)=\sum_\epsilon \upe^{-\upi\epsilon(\lfloor a \rfloor+1)}\upe^{\upi\epsilon{\left[1- \frac{T}{\tau_\mu}\slashed{a}\right]}_+}\widetilde A_\ell(\epsilon),
\end{eqnarray}
where $a\equiv s/T=\lfloor a\rfloor+\slashed{a}$, $\epsilon\equiv \omega T$ [$\ll 1$; see Eq.~\eqref{cond:shortT}], and $\widetilde A_\ell(\epsilon)=A_\ell(\omega=\epsilon/T)$. Note that $\tau_\Tar/T=1$, and we let $R\equiv\tau_g/T=\tau_\Sim/T \leq 1$. Putting all these into $\Lambda_{k\ell}(b)$, straightforward algebra yields
\begin{eqnarray}\label{eq:Lambda}
&&\Lambda_{k\ell}(b)\\
&=&\sum_\epsilon \widetilde A_\ell(\epsilon){\left[\sum_{q=0}^{p-1}\upe^{-\upi\epsilon(q+1)} I_{k\ell;q}(b;1)
+\upe^{-\upi\epsilon(p+1)} I_{k\ell;p}(b;\slashed{b}\,)\right]},\nonumber
\end{eqnarray}
where $p\equiv \lfloor b\rfloor$, $\slashed{b}= b-p\in[0,T)$, and
\begin{equation}\label{eq:Iklq}
I_{k\ell;q}(b;c)\equiv \int_0^c\!\!\!\upd a\,\widetilde f_{k\ell}(b-q- a){\left[\upe^{\upi\epsilon(1- a)}-\upe^{\upi\epsilon{\left[1- a/R\right]}_+}\!\right]}\,.
\end{equation}
Here, when $p=0$ so that $\sum_{q=0}^{p-1}$ seems to go from $0$ to $-1$, that sum is understood to be zero, so that only the second term within the brackets in Eq.~\eqref{eq:Lambda} is present.
Similarly, we have
\begin{eqnarray}
&&\overline{\Lambda}_{k\ell}(b)\\
&=&\sum_\epsilon \widetilde A_\ell(\epsilon){\left[\sum_{q=p}^{N-1}\upe^{-\upi\epsilon(q+1)} I_{k\ell;q}(b;1)
-\upe^{-\upi\epsilon(p+1)} I_{k\ell;p}(b;\slashed{b}\,)\right]}.\nonumber
\end{eqnarray}
Putting in the spectral function in place of $\widetilde f$, Eq.~\eqref{eq:Iklq} becomes
\begin{equation}
I_{k\ell;q}(b;c)=\int_{-\infty}^\infty\!\!\upd x \, \widetilde J_{k\ell}(x)\upe^{-\upi x(b-q)}D(c;x).
\end{equation}
where\\[-4ex]
\begin{equation}\label{eq:Dcx}
D(c;x)\equiv \int_0^c\upd a\,\upe^{\upi x a}{\left[\upe^{\upi \epsilon (1-a)}-\upe^{\upi \epsilon[1-a/R]_+}\right]}.
\end{equation}
Here, we assume that the $x$ and $a$ integrals are interchangeable, given regularity properties of $J_{kl}$. 

As we will evaluate the integral $I_{k\ell;q}$ above for $c\leq 1$, it depends on $\widetilde f_{kl}(a)$ only for $a\in[0,1]$, i.e., within a single stroboscopic time period $[0,T]$, for which there is no a priori reason for $I_{k\ell;q}$ to be small. Consequently, the $\Lambda$ functions generally need not be small. Thus even for $\mathscr{S}_1$, the dynamics of the simulator and target need not be close to each other. 

Below, we examine the $\Lambda$ functions in different parameter regimes.
For analytical estimates, it is simpler to consider $R=\tau_g/T$ is in two extreme regimes: $R\rightarrow0$ or $R\rightarrow 1$. The former corresponds to the common situation where the gate-pulse time is the shortest timescale in the problem; the latter can be thought of as a stroboscopic simulation scheme where the gate pulse is done as frequently as possible. For our single-gate exact simulator $\sS_1$, since $\HSsim$ is but a rescaled version of  $\HStar$, the $R\rightarrow 1$ regime gives $\HSsim=\HStar$ and $D(c;x)$---and consequently the $\Lambda$ functions---vanishes. Thus, only the regime of $R\rightarrow 0$ is nontrivial for $\sS_1$. In the remainder of the paper, whenever $\mathscr{S}_1$ occurs, $R$ is taken to approach 0, in which case, $D(c;x)\vert_{R\rightarrow 0}\equiv D_0(c;x)$ can be worked out exactly:
\begin{equation}\label{eq:D0}
D_0(c;x)=\frac{\upi{\left[\epsilon{\left(1\!-\!\upe^{\upi xc}\right)}-x{\left(1\!-\!\upe^{\upi \epsilon}\right)}+\upe^{\upi xc}x{\left(1\!-\!\upe^{\upi\epsilon(1-c)}\right)}\right]}}{x(x-\epsilon)}.
\end{equation}

We consider three parameter regimes amenable to analytical estimates (we look outside of these regimes in the numerical analysis of Sec.~\ref{sec:numerics}):
\begin{eqnarray*}
\textrm{regime I}:&\qquad& |\bar x|,x_c\ll\epmax\ll 1;\\
\textrm{regime II}:&\qquad&\epmax\ll |\bar x|,x_c\ll 1;\\
\textrm{regime III}:&\qquad&\epmax\ll 1\ll |\bar x|, x_c.
\end{eqnarray*}
Here, $\epmax\equiv \max |\epsilon|=\ommax T$, where $\ommax$ is the largest transition frequency for the target system. $|\bar x|, x_c\ll \epmax$ (regime I) or $\ll 1$ (regime II) means that all relevant values of $x$ are such that $|x|\ll \epmax$ or $\ll 1$. Similarly, $|\bar x|, x_c\gg 1$ (regime III) tells us that $|x|\gg 1$ is the domain of interest.
Appendix \ref{app:D0} shows that $D_0(c;x)$ in these three regimes can be approximated as
\begin{equation}
D_0(c;x)\simeq{\left\{\begin{array}{ll}
\upi\epsilon c{\left(1-\tfrac{c}{2}\right)}&\quad\textrm{for regimes I \& II}
\\[0.5ex]
-\frac{\epsilon}{x}[1-(1-c)\upe^{\upi xc}]&\quad\textrm{for regime III}
\end{array}
\right.}.
\end{equation}
In the following subsections, we calculate the simulation error $\tfrac{1}{\alpha^2}\Err{N}$ for the different regimes and discuss the physical implications.

\subsubsection{Regimes I \& II}
In regimes I and II, the $\Delta_i(\cdot)$s can be approximated as (see Appendix \ref{app:Deltai}),
\begin{eqnarray}\label{eq:delta1_maintext}
\Delta_1(\cdot)&\simeq&\tfrac{\upi}{2} N^2\sum_{k\ell}\widetilde f_{k\ell}(0)\sum_{\epsilon\epsilon'}(\epsilon+\epsilon')\widetilde A_\ell(\epsilon)(\cdot)\widetilde A_k(\epsilon'),\\
\label{eq:delta2_maintext}\Delta_2(\cdot)&\simeq&-\tfrac{\upi}{4}N^2\!\sum_{k\ell}\sum_{\epsilon\epsilon'}\widetilde{A}_k(\epsilon')\widetilde{A}_\ell(\epsilon)(\cdot)\widetilde{f}_{k\ell}(0)(\epsilon\!+\!\epsilon')\\
\Delta_3 (\cdot)&=&[\Delta_2(\cdot)]^\dagger\nonumber
\end{eqnarray}
The $\widetilde A$ operators, by definition, have norm of order unity. We thus see that, in regimes I and II, $\frac{1}{\alpha^2}\Err{N}=\Vert\Delta_1+\Delta_2+\Delta_3\Vert$ is approximately (up to a constant factor that depends on the choice of the norm and corrections higher-order in small quantities)
\begin{equation}\label{cond12}
\tfrac{1}{\alpha^2}\Err{N}\sim N^2\epmax\widetilde{f}(0)=(NT)^2(\omega_{\max}T)f(0),
\end{equation}
where $\widetilde f(0)\equiv \max_{k\ell}|\widetilde f(0)|\equiv T^2 f(0)$, and the $O(N)$ terms are treated as subdominant. Here, we have assumed that $N$ is such that $N\epmax, N|\bar x|, Nx_c\ll 1$.

The requirement that Eq.~\eqref{cond12} is small, together with the regime conditions of $|\bar x|,x_c\ll\epmax\ll 1$ (regime I) or $\epmax\ll |\bar x|,x_c\ll 1$ (regime II), gives the criteria under which the open-system dynamics of the simulator $\mathscr{S}_1$ and that of the target are stroboscopically close to each other for (at least) $N$ cycles. That the error $\Err{N}$ grows with $N^2$ [or time $(NT)^2$] comes from the second-order perturbation theory. It is plausible that a different approach to the analysis might yield a different dependence on $N$, but we do not expect that dependence to disappear: The simulation error will accumulate as time passes. 

Regime I, with $x_c\ll \epmax \ll 1$, when translated to dimensional quantities, entails the condition $\nu_c=1/\tauB\ll \ommax \ll 1/T$. This requires $\ommax^{-1}\ll\tauB$ as well as $T\ll \tauB$. The requirement of $\ommax^{-1}\ll\tauB$ puts a restriction on the target Hamiltonian: The  intrinsic target timescales must be much shorter than the bath correlation time $\tauB$. This is typically the regime of non-Markovian dynamics on the system~\cite{Breuer:2002:OxfordUniversityPress}. The requirement of $T\ll \tauB$ suggests that the coarse-graining in time according to $T$, introduced by the stroboscopic simulation cycles, is not ``visible'' to the bath---it sees only the effective stroboscopic dynamics, and has no time to respond to fast changes in $\sS_1$ occuring within time $T$. The bath thus sees the simulator dynamics as close to that of the target, and the open-system simulation error is small \cite{lloyd1999robust}.

For regime II, with $\epmax\ll |\bar x|,x_c\ll 1$, one has instead $\ommax \ll \nu_c= 1/\tauB \ll T$, so that $\ommax^{-1} \gg \tauB$ and $T\ll \tauB$. In this case, the target timescales are much longer than the correlation time of the bath, as is typical for Markovian dynamics~\cite{Breuer:2002:OxfordUniversityPress}. Even so, as long as $T$ is much smaller than $\tauB$, the bath is still unable to react to the fast changes of $\mathscr{S}_1$. However, $\tauB$ is typically small in most situations, so $T$ must be extremely short in order for this regime to apply, which may be unattainable in practice.

In Eq.~\eqref{cond12}, $NT$ should be regarded as the total simulation time $t$. If we keep $t$ fixed (i.e., changing $N$ as $T$ changes such that $t$ is constant), then, the simulation error scales linearly with $T$, provided, of course, that the conditions for regimes I and II remain valid as $T$ changes for the above analysis to hold. Note that $\omega_{\max}$ and $f(0)$ are quantities having to do with the target Hamiltonian, and with the bath; both do not change as $T$ changes. Thus, if $T$ is shortened, the simulation error decreases.

\subsubsection{Regime III}\label{sec:regime3}
For regime III, $\widetilde J(x)$ is significant only when $|x|$ is large, and the $\Delta_i$s can be examined in this limit.
In addition, the assumption of $x_c\gg 1$ for regime III says that the width of $\widetilde{f}_{k\ell}(a)$, $a_B\equiv 1/x_c= \tauB/T$, is much less than 1, so $\widetilde{f}_{k\ell}(a)$ is negligible whenever $|a|>1$. Then $I_{k\ell;q}(b;c)$ is negligible except when $q= \lfloor b\rfloor$ or $\lfloor b\rfloor\pm 1$. With these approximations, we show in Appendix~\ref{app:Deltai} that $\Delta_1$ is insignificant compared to 
$\Delta_2$ and $\Delta_3$, and that $\Delta_2$ is given by
\begin{equation}\label{eq:D23III}
\Delta_2(\,\cdot\,) \simeq \frac{\upi}{2} N\!\sum_{k\ell}\!\sum_{\epsilon\epsilon'}\widetilde{A}_{k}(\epsilon')\widetilde{A}_\ell(\epsilon)(\,\cdot\,)(\epsilon+\epsilon')\!\!\!\int_{-\infty}^0\!\!\!\!\upd a \widetilde{f}_{k\ell}(a;x_c);
\end{equation}
$\Delta_3(\,\cdot\,)=\Delta_2(\,\cdot\,)^\dagger$ yields the approximation for $\Delta_3$. The linear $N$-dependence of $\Delta_2$ and $\Delta_3$ here, instead of the quadratic dependence for regimes I and II, can be understood as follows: The two factors of $N$ in regimes I and II came from, first, the integral over $b$ from $0$ to $N$, and second, the sum over $q$ from $0$ to $N$ in the $\Delta_i$s. In regime III, as argued above, the sum over $q$ is reduced to a sum over the three possible values of $\lfloor b\rfloor$, $\lfloor b\rfloor +1$ and $\lfloor b\rfloor -1$, independent of $N$. The remaining integral over $b$ from $0$ and $N$ gives the factor of $N$ in Eq.~\eqref{eq:D23III}.

Since $a_B\ll 1$, we estimate
\begin{eqnarray}\label{cond3}
\tfrac{1}{\alpha^2}\Err{N}\sim ~&&N\epmax a_B\sup_{a\in \mathbb{R}}|\widetilde{f}(a)|\\
&&=(NT)(\ommax T)(\tauB/T)T\sup_{s\in \mathbb{R}}|f(s)|.\nonumber
\end{eqnarray}
The requirement that Eq.~\eqref{cond3} is small, together with the regime III conditions of $\epmax\ll 1\ll |\bar x|, x_c$, gives criteria for the close simulation of the target.
Regime III assumes $\epmax \ll x_c$, or, equivalently, $\ommax^{-1}\gg \tauB$. This is also the regime of Markovian system dynamics in the weak-coupling limit. Furthermore, we have $1/x_c \ll 1$, which means $T \gg \tauB$. Unlike regimes I and II, here, $\sS_1$ can be stroboscopically close to the target even when $T$ is much larger than the correlation time of the bath, as long as Eq.~\eqref{cond3} is small. This is a surprising result, and contrary to the requirement of $T\ll\tauB$ standard in past quantum simulator discussions: Even if the bath has sufficient time to respond to a \emph{slow} change in $\sS_1$, good simulation is still possible.

Here, $NT$ should again be regarded as the total simulation time $t$. If we consider $t$ fixed, the simulation error in regime III, as in regimes I and II, scales linearly with $T$. Thus, as before, one can reduce the simulation error by decreasing $T$.

Note that the analysis above gives sufficient conditions for close simulation. There is no \emph{a priori} assumption of Markovian dynamics, a feature often imposed in past discussion of this question.

\subsection{Zero-temperature oscillator bath}\label{sec:0_temp_ohmic_bath}

We now examine an analytically tractable system-bath model, which serves as an additional check on the conditions of the last subsection. Consider a system coupled to a bath of harmonic oscillators (a bosonic bath), with the bath Hamiltonian
\begin{equation}
\HB = \sum_m\omega_m b_m^\dagger b_m,
\end{equation}
and the system-bath coupling
\begin{equation}
\HSB = \sum_k A_k \otimes B_k,~~\textrm{with }B_k\equiv\sum_m (g_{km} b_m + g_{km}^* b_m^\dagger).
\end{equation}
Here, $b_m$ and $b_m^\dagger$ are annihilation and creation operators for mode $m$, satisfying the bosonic commutation relations of $[b_m,b_n]=0=[b_m^\dagger,b_n^\dagger]$ and $[b_m,b_n^\dagger]=\delta_{mn}$. $g_{km}$ is the system-bath coupling constant for mode $m$, for the system operator $A_k$. 

Suppose the bath is initially in the $\HB$-thermal state $\rhoB=\upe^{-\beta \HB}/Z$, where $\beta$ is the inverse temperature, and $Z\equiv\Tr\left(\upe^{-\beta \HB}\right)$ is the partition function. The bath correlation function is then
\begin{eqnarray}
f_{k\ell} (s)&=&\Tr_\mathrm{B}\{B_k(s)B_\ell\rho_B\}\\
&=&\sum_mg_{km}^*g_{\ell m}{\left\{[1+\cN(\omega_m)] \upe^{-\upi\omega_m s} + \cN(\omega_m) \upe^{\upi\omega_ms}\right\}}\nonumber\\
&=&\int_{-\infty}^\infty \upd\nu\, \cJ_{k\ell}(\nu){\left\{[1+\cN(\nu)]\upe^{-\upi\nu s}+\cN(\nu)\upe^{\upi\nu s}\right\}}.\nonumber
\end{eqnarray}
Here, $\cN(\omega_m)=\Tr_\mathrm{B}(b_m^\dag b_m\rho_B)$
is the average particle-number for mode $m$. For the thermal bath state,
$\cN(\omega_m)= 1/{\left(\upe^{\beta \omega_m}-1\right)}$.
In addition, we have done the standard replacement of the discrete sum $\sum_m g_{km}^*g_{\ell m}\{\cdot\}$ by the continuous integral $\int_{-\infty}^\infty \upd\nu\cJ_{k\ell}(\nu) \{\cdot\} $, where $\cJ_{k\ell}(\nu)$ is the spectral density of the bath. 
A commonly used form for the spectral density is
\begin{equation}\label{eq:OhmicSpec}
\cJ_{k\ell}(\nu)=
{\left\{\begin{array}{ll}
\eta_{k\ell} \,\nu^w\,\upe^{-\nu/\nu_{c;k\ell}} &\textrm{for }\nu \geq 0 \\[0.5ex]
0  &\textrm{for }\nu < 0
\end{array}\right.},
\end{equation}
where $\nu_{c;k\ell}$ is the cutoff frequency. For simplicity, we set $\nu_{c;k\ell}=\nu_c$ ($x_{c;k\ell}=x_c$), and $\eta_{ k\ell}=\eta\delta_{k\ell}$, for all $k, \ell$, where $\delta_{k\ell}$ is the Kronecker delta, and $\eta$ is a real constant. Note that $\eta$ here plays the role of the small book-keeping parameter $\alpha$ of the earlier analysis. $\cJ$ is often referred to as Ohmic when $w=1$, sub-Ohmic when $w<1$, and super-Ohmic when $w>1$.

Consider the case of zero temperature, for which the bath state is $\rhoB=|0\rangle\langle0|$, and $\cN(\nu)=0$. In this case, $\cJ_{k\ell}(\nu)$ is the Fourier transformation of $f_{k\ell}(s)$, i.e., $\cJ_{k\ell}(\nu)=J_{k\ell}(\nu)$, or
\begin{equation}\label{eq:ohmicJ}
\widetilde J_{k\ell}(x)=TJ_{k\ell}{\left(\nu=\frac{x}{T}\right)}={\left\{\begin{array}{ll}\widetilde\eta\,\delta_{k\ell}\,x^w\upe^{-x/x_c},&x\geq 0\\[0.5ex]
0,  &x < 0
\end{array}\right.},
\end{equation}
for the dimensionless version, where $\widetilde\eta\equiv \eta T^{1-w}$, and $x_c\equiv \nu_cT$.

Let us estimate the size of $\Delta_i$s for the situation of the zero-temperature Ohmic bath for the different parameter regimes. Note that the Ohmic bath cannot be considered in regime III, for which the domain of interest is $|x|\gg 1$ as the Ohmic bath $\widetilde J$ has significant support near $x=0$. We thus content ourselves with only regimes I and II. 

For $\Delta_1$, starting from Eq.~\eqref{eq:Delta1}, in the limit of regimes I and II, we have
\begin{eqnarray}
\Delta_1(\,\cdot\,)&\simeq&N^2\!\sum_{k\ell}\!\sum_{\epsilon\epsilon'}{\left[\widetilde A_\ell(\epsilon)(\cdot)\widetilde A_k(\epsilon')\!\!\int_{-\infty}^\infty\!\!\upd x\widetilde J_{k\ell}(x)D_0(1;x)\right.}\nonumber\\
&&{\left.+\widetilde A_k(\epsilon')(\cdot)\widetilde A_\ell(-\epsilon)\!\!\int_{-\infty}^\infty\!\!\upd x\widetilde J_{k\ell}(x)^*D_0(1;x)^*\right]}.~
\end{eqnarray}
Now, putting in $\widetilde J$ for the Ohmic bath, and from the definition of $D_0$ [Eq.~\eqref{eq:Dcx}], we have
\begin{eqnarray}\label{eq:JD}
\int_{-\infty}^\infty\!\!\upd x\widetilde J_{k\ell}(x)D_0(1;x)&=&\delta_{k\ell}\,\widetilde\eta \,x_c^2\!\!\int_0^1\!\!\!\upd a\frac{\upe^{\upi\epsilon(1-a)}-1}{(1-\upi ax_c)^2}\\
&\simeq&\delta_{k\ell}\,\widetilde\eta \,x_c^2\!\!\int_0^1\!\!\!\upd a\frac{\upi\epsilon(1-a)}{(1-\upi ax_c)^2}\nonumber\\
&=&\delta_{kl}\,\widetilde\eta\,\epsilon \,{\left[-x_c+\upi \ln(1-\upi x_c)\right]}.\nonumber
\end{eqnarray}
In regimes I and II, we can expand $\ln (1-\upi x_c)$ to second order in $x_c$, and approximate the above integral by $\tfrac{\upi}{2} \delta_{k\ell}\widetilde\eta\epsilon x_c^2$. Then,
\begin{eqnarray}
\Delta_1(\,\cdot\,)&\simeq&N^2\sum_{k\ell}\sum_{\epsilon\epsilon'}\widetilde A_\ell(\epsilon)(\cdot)\widetilde A_k(\epsilon')\tfrac{\upi}{2}\delta_{k\ell}\,\widetilde\eta\,x_c^2(\epsilon+\epsilon'),~~
\end{eqnarray}
which we observe to be exactly the expression in Eq.~(\ref{eq:delta1_maintext}) for $\Delta_1(\cdot)$, upon noting that $\widetilde f_{k\ell}(0)=\int_{-\infty}^\infty\upd x\widetilde J_{k\ell}(x)=\delta_{k\ell}\widetilde\eta x_c^2$ for the zero-temperature Ohmic bath.

For $\Delta_2(\cdot)$, Eq.~\eqref{eq:Delta2} gives, in the regimes of I and II,
\begin{eqnarray}
&&~\Delta_2(\,\cdot\,)\\
&\simeq&-N\sum_{k\ell}\sum_{\epsilon\epsilon'}{\left[\widetilde A_k(\epsilon')\widetilde A_\ell(\epsilon)\!\!\int_{-\infty}^\infty\!\! \upd x\widetilde J_{k\ell}(x)D_0(1;x)\tfrac{N-1}{2}\right.}\nonumber\\
&&~+\widetilde A_\ell(\epsilon)\widetilde A_k(\epsilon')\!\!\int_{-\infty}^\infty\!\! \upd x\widetilde J_{k\ell}(x)^*D_0(1;-x)\tfrac{N+1}{2}\nonumber\\
&&~+\widetilde A_k(\epsilon')\widetilde A_\ell(\epsilon)\!\!\int_{-\infty}^\infty\!\! \upd x\widetilde J_{k\ell}(x)\!\!\int_0^1\!\!\upd b'D_0(b';x)\upe^{-\upi(x+\epsilon')b'}\nonumber\\
&&~{\left.-\widetilde A_\ell(\epsilon)\widetilde A_k(\epsilon')\!\!\int_{-\infty}^\infty\!\! \upd x\widetilde J_{k\ell}(x)^*\!\!\!\int_0^1\!\!\upd b'D_0(b';-x)\upe^{-\upi xb'}\right]}(\cdot).\nonumber
\end{eqnarray}
As in Eq.~\eqref{eq:JD}, we have $\int_{-\infty}^\infty\upd x\widetilde J_{k\ell}(x)D_0(1;\pm x)\simeq \delta_{k\ell}\,\widetilde\eta\,\epsilon\,[\mp x_c+\upi\ln(1\mp\upi x_c)]\simeq\tfrac{\upi}{2}\delta_{k\ell}\,\widetilde\eta\,\epsilon\, x_c^2$. Note that $\widetilde J_{k\ell}(x)^*=\widetilde J_{k\ell}(x)$. In addition, we need the following integral,
\begin{eqnarray}
&&\int_{-\infty}^\infty\!\! \upd x\widetilde J_{k\ell}(x)\!\!\int_0^1\!\!\upd b'D_0(b';x)\upe^{-\upi(x+\epsilon')b'}\\
&=&\delta_{k\ell}\widetilde \eta\!\int_0^1\!\!\upd b'\upe^{-\upi\epsilon' b'}\!\!\!\!\int_0^{b'}\!\!\!\upd a{\left[\upe^{\upi\epsilon (1-a)}-1\right]}\!\!\int_0^\infty\!\!\!\upd x \,x\upe^{-x/x_c}\upe^{\upi x(a-b')}\nonumber\\
&\simeq&\delta_{k\ell}\widetilde\eta\frac{\epsilon}{2 x_c}{\left[x_c(2\upi+x_c)-2\ln(1+\upi x_c)\right]}\simeq\delta_{k\ell}\,\widetilde\eta\,\frac{\upi \epsilon\, x_c^2}{3},\nonumber
\end{eqnarray}
since \mbox{$\upe^{\upi\epsilon(1-a)}-1\simeq\upi\epsilon (1-a)$} and $\upe^{-\upi\epsilon' b'}\simeq 1$. Similarly, $\int_{-\infty}^\infty\! \upd x\widetilde J_{k\ell}(x)^*\int_0^1\!\upd b'D_0(b';-x)\upe^{-\upi xb'}\simeq\delta_{k\ell}\widetilde\eta\,\frac{\upi \epsilon x_c^2}{3}$. Consequently,
\begin{eqnarray}
&&~\Delta_2(\,\cdot\,)\\
&\simeq&-\upi N\delta_{k\ell}\widetilde\eta x_c^2\nonumber\\
&&\quad \times \sum_{k\ell}\sum_{\epsilon\epsilon'}{\left[\widetilde A_k(\epsilon')\widetilde A_\ell(\epsilon)\epsilon \tfrac{N-1}{4}+\widetilde A_\ell(\epsilon)\widetilde A_k(\epsilon')\epsilon \tfrac{N+1}{4}\right.}\nonumber\\
&&{\left.\qquad\qquad\qquad+\widetilde A_k(\epsilon')\widetilde A_\ell(\epsilon)\,\tfrac{\epsilon}{3}-\widetilde A_\ell(\epsilon)\widetilde A_k(\epsilon')\,\tfrac{\epsilon}{3}\right]}(\cdot)\nonumber\\
&=&-\upi N\delta_{k\ell}\widetilde\eta x_c^2\sum_{k\ell}\sum_{\epsilon\epsilon'}\widetilde A_k(\epsilon')\widetilde A_\ell(\epsilon)\nonumber\\
&&\qquad\qquad\qquad\times {\left[\tfrac{N}{4}(\epsilon+\epsilon')+\tfrac{1}{12}(\epsilon-\epsilon')\right]}(\cdot)\nonumber,
\end{eqnarray}
which is exactly Eq.~\eqref{eq:delta2_maintext} upon retaining only the $O(N^2)$ term.

\subsection{Multi-gate Exact Simulator}\label{sec:MgateDQS}
Practically, one expects to use multiple gates---not just a single gate as in $\sS_1$---per simulation cycle to achieve good simulation of the target Hamiltonian. As before, we assume that the $M$-gate simulator, $\sS_M$, is exact, so that $g_M\ldots g_2g_1=\exp(-\upi\HStar T)$. Here, the gate $g_m$ is assumed to be applied instantaneously at time $t_m$. The last gate in the simulation cycle $g_M$ occurs at time $t_M\equiv \tau_g+t_1$, after which no gates are applied until the next cycle begins (see Fig.~\ref{fig:gate_sequence}). The sequence of $M$ gates is thus completed in time $\tau_g$, and we define $R_M\equiv\tau_g/T$ to denote the fraction of $T$ taken for the $M$ gates to be applied in each cycle. (Note: $R_1$ is exactly the $R$ quantity for $\sS_1$ of Sec.~\ref{sec:1gateDQS}; there, we took $R_1\rightarrow 0$ for instantaneous gates.)

As in the case of $\sS_1$, we want to bound the $N$-cycle simulation error $\tfrac{1}{\alpha^2}\Err{N}$, comparing $\sS_M$ to the target under different parameter regimes. It is convenient to split $\Err{N}$ into two pieces,
\begin{eqnarray}
\Err{N}&=&\Vert\cE_{N,I}^{(\Tar)}-\cE_{N,I}^{(\sS_M)}\Vert\equiv \Err{N}(\Tar,\sS_M)\nonumber\\
&\leq& \underbrace{\Vert\cE_{N,I}^{(\Tar)}-\cE_{N,I}^{(\sS_1)}\Vert}_{\Err{N}(\Tar,\sS_1)}+\underbrace{\Vert\cE_{N,I}^{(\sS_1)}-\cE_{N,I}^{(\sS_M)}\Vert}_{\Err{N}(\sS_1,\sS_M)}.
\end{eqnarray}
$\Err{N}(\Tar,\sS_1)$ is the simulation error between $\sS_1$ and the target, which we already analyzed in Sec.~\ref{sec:1gateDQS}; the second piece $\Err{N}(\sS_1,\sS_M)$ compares $\sS_1$ to $\sS_M$, which we bound below. By splitting the simulation error between $\sS_M$ and the target into these two pieces, we analyse the errors due to the stroboscopicity [captured by $\Err{N}(\Tar,\sS_1)$] and the multiple gates [captured by $\Err{N}(\sS_1,\sS_M)$] separately. The $\sS_1$ considered here---artificially inserted to help bound $\Err{N}$---has a single gate, generated by a Hamiltonian $\propto \HStar$, that lasts for time no longer than $\tau_g=t_M$ (in the limit we are considering here, that single gate is in fact instantaneous). This means that we have
\begin{eqnarray}\label{eq:US1USM}
U_{\sS_1}(NT+t,NT)&=&U_{\sS_M}(NT+t,NT)\\
&=&\UStar((N+1)T,NT)=\upe^{-\upi\HStar T}.\nonumber
\end{eqnarray}
for $t\in[\tau_g, T]$, or, equivalently, that
\begin{equation}
U_{\sS_1}\!\bigl((N\!+\!1)T,NT+t\bigr)\!=\!U_{\sS_M}\!\bigl((N\!+\!1)T,NT+t\bigr)\!=\!\mathbbm{1}. 
\end{equation}

\begin{table*}
\begin{tabular}{c|c|c|c}
\hline
\hline
Regime & Conditions & Bound on the simulation error $\tfrac{1}{\alpha^2}\Err{N}$&Equations\\[0.8ex]
\hline
\rule[1.5ex]{0pt}{8pt}I & $\quad|\bar x|,x_c\ll\epmax\ll 1\quad$& $N^2{\left[C\epmax\widetilde f(0)+8R_M\sup_{a\geq 0}|\widetilde f(a)|\right]}\qquad\qquad$&\multirow{2}{*}{\eqref{cond12} and \eqref{cond1M}}\\
\cline{1-2}
II & $\epmax\ll |\bar x|,x_c\ll 1$&$\qquad=(NT)^2{\left[C(\ommax T)f(0)+8(\tau_g/T)\sup_{s\geq 0}|f(s)|\right]}\qquad$&\\[0.8ex]
\hline
\rule[1.5ex]{0pt}{8pt}\multirow{2}{*}{III} & \multirow{2}{*}{$\epmax\ll 1\ll |\bar x|, x_c$}& $N^2{\left[C\tfrac{1}{N}\epmax a_B\sup_{a\in \mathbb{R}}|\widetilde{f}(a)|+8R_M\sup_{a\geq 0}|\widetilde f(a)|\right]}$&\multirow{2}{*}{\eqref{cond3} and \eqref{cond1M}}\\
&&$\qquad=(NT)^2{\left[C\tfrac{1}{N}(\ommax T)(\tauB/T)\sup_{s\in \mathbb{R}}|f(s)|+8(\tau_g/T)\sup_{s\geq 0}|f(s)|\right]}\qquad$&\\[0.8ex]
\hline
\hline
\end{tabular}
\caption{\label{tab:condition} Summary of the simulation errors under the different regimes. Stated above are the bounds for $\sS_M$, up to an overall constant. To recover the bounds for $\sS_1$ from Sec.~\ref{sec:1gateDQS}, set $R_{M=1}=0$. $C$ is the constant $C\equiv \sum_{\epsilon\epsilon'}1$, giving the relative factor between $\Err{N}(\Tar,\sS_1)$ and $\Err(\sS_1,\sS_M)$ coming from the $\epsilon$ and $\epsilon'$ sums in Eqs.~\eqref{cond12} and \eqref{cond3}, absent from Eq.~\eqref{cond1M}.}
\end{table*}

Comparing $\sS_M$ to $\sS_1$, we have, from Eq.~\eqref{eq:Lambdas},
\begin{eqnarray}
&&\Lambda_{k\ell}(b)\equiv \Lambda_{k\ell}^{(\sS_1,\sS_M)}(b)\\
&=&\int_0^b\!\!\upd a~\widetilde f_{k\ell}(b-a){\left[\widetilde A_\ell^{(\sS_1)}(a)-\widetilde A_\ell^{(\sS_M)}(a)\right]}\nonumber\\
&=&\!\!\sum_{q=0}^{\lfloor b\rfloor -1}\!\!\!\int_0^{R_M}\!\!\!\upd a\widetilde f_{k\ell}\bigl(b\!-\!(q\!+\!a)\bigr){\left[\widetilde A_\ell^{(\!\sS_1\!)}(q\!+\!a)\!-\!\widetilde A_\ell^{(\!\sS_M\!)}(q\!+\!a)\right]}\nonumber\\
&+&\!\!\int_0^{\min\{\!\slashed{b},R_M\!\}}\!\!\!\!\upd a\widetilde f_{k\ell}(\slashed{b}\!-\!a){\left[\widetilde A_\ell^{(\!\sS_1\!)}\!(\lfloor b\rfloor\!+\!a)\!-\!\widetilde A_\ell^{(\!\sS_M\!)}\!(\lfloor b\rfloor\!+\!a)\right]},\nonumber
\end{eqnarray}
where we have used Eq.~\eqref{eq:US1USM}, with $\widetilde A_\ell^{(\cdot)}(a)=U_\cdot(aT)^\dagger A_\ell U_\cdot(aT)$, to infer that $\widetilde A_\ell^{(\sS_1)}\bigl((q+a)T\bigr)-\widetilde A_\ell^{(\sS_M)}\bigl((q+a)T\bigr)=0$ for $q$ a nonnegative integer and $a\in [R_M,1]$, so that the upper limits of the $a$ integrals read as given above. The expressions in the brackets $[\ldots ]$, unlike in the comparison of $\sS_1$ and the target, are generally complicated. However, they can be straightforwardly bounded as, for any $a$,
\begin{eqnarray}
&&{\left\Vert\widetilde A_\ell^{(\sS_1)}(a)-\widetilde A_\ell^{(\sS_M)}(a)\right\Vert}\\
&\leq& {\left\Vert\widetilde A_\ell^{(\sS_1)}(a)\right\Vert}+{\left\Vert\widetilde A_\ell^{(\sS_M)}(a)\right\Vert}=2\Vert A_\ell\Vert,\nonumber
\end{eqnarray}
for a unitarily invariant norm. 
Thus, we have,
\begin{equation}
{\left\Vert\Lambda_{k\ell}^{(\sS_1,\sS_M)}(b)\right\Vert}\leq 2\sup_{a\in[0,b]}|\widetilde f_{k\ell}(a)|\bigl(\lfloor b\rfloor+1\bigr)R_M\Vert A_\ell\Vert.
\end{equation}
A similar analysis gives the bound for $\overline\Lambda_{k\ell}(b)$:
\begin{equation}
{\left\Vert \overline\Lambda_{k\ell}^{(\sS_1,\sS_M)}(b)\right\Vert}\leq 2\sup_{a\in[0,b]}|\widetilde f_{k\ell}(a)|\bigl(N-\lfloor b\rfloor+1\bigr)R_M\Vert A_\ell\Vert.
\end{equation}
Using these bounds on $\Lambda_{k\ell}(b)$ and $\overline\Lambda_{k\ell}(b)$ in Eq.~\eqref{eq:Delta12_original}, we have
\begin{eqnarray}
&&\tfrac{1}{\alpha^2}\Err{N}(\sS_1,\sS_M)\\
&\leq &8\sum_{k\ell}\Vert A_k\Vert\Bigl[2\sup_{a\in[0,N]}|\widetilde f_{k\ell}(a)|R_M\Vert A_\ell\Vert\Bigr]\int_0^N\!\!\!\upd b(\lfloor b\rfloor+1)\nonumber\\
&=&8N(N+1)R_M\sum_{k\ell}\sup_{a\in[0,N]}|\widetilde f_{k\ell}(a)|\Vert A_k\Vert\Vert A_\ell\Vert.\nonumber
\end{eqnarray}
Hence,
\begin{eqnarray}\label{cond1M}
\tfrac{1}{\alpha^2}\Err{N}(\sS_1,\sS_M)&\sim& N^2R_M\sup_{a\geq 0}|\widetilde{f}(a)|\\
&=&(NT)^2(\tau_g/T)\sup_{s\geq 0}|f(s)|.\nonumber
\end{eqnarray}
The difference between $\sS_1$ and $\sS_M$ vanishes as $R_M\rightarrow 0$, as can be expected: That there are $M$ gates rather than one, in a time shorter than any timescales of the problem, cannot be physically relevant.

If we regard $NT$ as the simulation time $t$ to be kept fixed, the bound in Eq.~\eqref{cond1M} suggests that the $\tfrac{1}{\alpha^2}\Err{N}(\sS_1,\sS_M)$ piece of the simulation error follows an inverse relation to $T$, if we also hold $\tau_g$ constant.
This seems to indicate the possibility of reducing the simulation error by \emph{increasing} $T$. However, a larger $T$ means, at least in regimes I--III, larger $\frac{1}{\alpha^2}\Err{N}(\Tar,\sS_1)$. In the end, it is a balance between both pieces that will determine whether increasing or decreasing $T$ will reduce the total simulation error. In Sec.~\ref{sec:numerics} below, we see a specific example where the balance is such that a larger $T$ gives a smaller overall simulation error.

Table \ref{tab:condition} gathers the bounds for the simulation error under the different regimes analysed in Secs.~\ref{sec:1gateDQS} and \ref{sec:MgateDQS}.

\section{Numerical Examples}\label{sec:numerics}

To verify our analytical estimations of the previous section, and to explore the simulation efficacy beyond the regimes of I--III, in this section, we look at numerical studies of two target models. Specifically, we numerically calculate, as a function of time, the density matrix of the simulator exposed to open-system dynamics, and compare it with that of the target system.

\subsection{A toric-code vertex}

For feasible numerical computation in reasonable time, we consider a small system of four qubits, with the target Hamiltonian
\begin{equation}\label{eq:toric_H}
\HStar=-\frac{\omega}{2} X_1X_2X_3X_4,
\end{equation}
where $X_i$ is the Pauli $X$ operator acting on qubit $i$. The energy spectrum of this system is simple: There are two degenerate energy eigenspaces with eigenvalues $\{\omega/2,-\omega/2\}$, and there is a single transition frequency $\omega$. The ground space of $\HStar$ possesses a definite eigenvalue (of $+1$) of the operator $X_1X_2X_3X_4$, such that a $Z$ error on any one of the four qubits can be detected as a change in sign---since a single $Z$ anticommutes with $X_1X_2X_3X_4$---in the eigenvalue of the system state. That single $Z$ error excites the system from the ground state into the excited-state manifold, is energetically unfavorable given the $\omega$ gap, and hence is naturally suppressed.

The four qubits interacting under $\HStar$ can be thought of as a single-vertex piece of the Kitaev toric-code model \cite{Kitaev:2003:2}. The toric code is important in the context of fault-tolerant quantum computation~\cite{fowler2009highthreshold,Folwer2012PhysRevA.86.032324}, given its natural tolerance to local errors, its scalable structure, its high noise
threshold, and its instrinsic resilience against thermal noise~\cite{Dennis:2002:4452,Breuer:2002:OxfordUniversityPress,fowler2009highthreshold,Folwer2012PhysRevA.86.032324}. Recent experimental progress in this direction (for example,
see~\cite{barends_Martinis2014superconducting}) has further intensified interest in the model. The single-vertex example we study here, despite its small size, can still have relevance if the coupling to the bath is local and that the correlation legnth decays rapidly. Moreover, such codes are expected to provide noise good protection at large system sizes, so if a small system already demonstrates resilience to noise, one expects even better performance as the system scales up.

\begin{figure}
\centering\includegraphics[width=40mm]{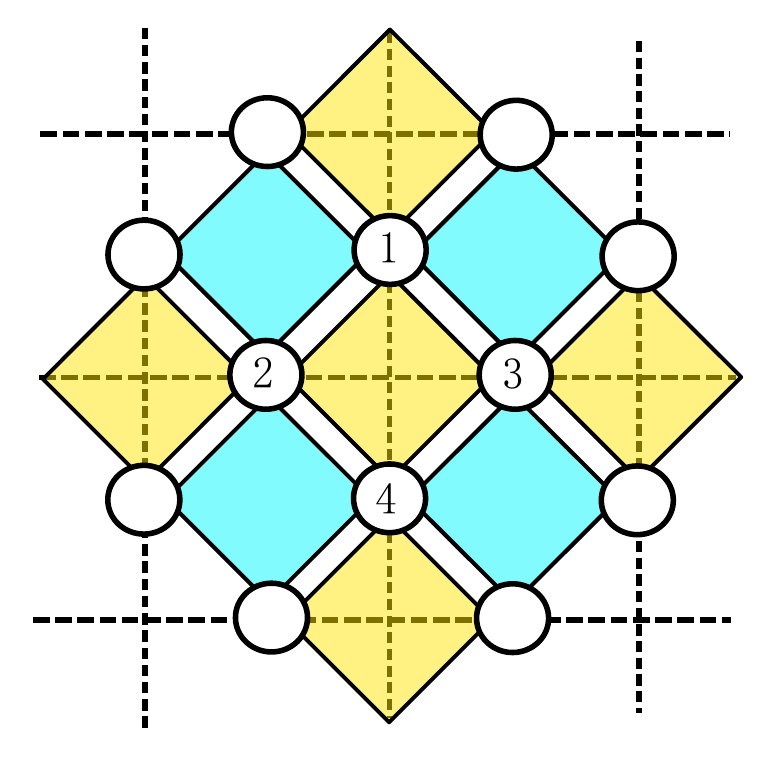}
\caption{\label{fig:simple_kitaev} A small piece of the square lattice in the toric-code model of \cite{Kitaev:2003:2}. Qubits (circles) sit on the edges of the lattice; qubits around each plaquette of the lattice (in cyan) are acted upon by the local four-body $ZZZZ$ terms of the toric-code model; the qubits around each lattice vertex (in yellow) are acted upon by the $XXXX$ terms. The qubits labeled 1 to 4 around the vertex are the four qubits in our $\HStar$ model of Eq.~\eqref{eq:toric_H}.}
\end{figure}

$\HStar$ is a four-body Hamiltonian---the full toric-code model also comprises four-body terms (see Fig.~\ref{fig:simple_kitaev})---which is typically difficult to engineer in the lab. Instead, one approach to achieve a toric-code interaction is to make use of DQS, and decompose the desired target Hamiltonian into a sequence of gates \cite{RydbergsimulatorNatPhysics,young2012finitetemperature_toric,becker2013dynamic_self_correction}.
For our four-qubit situation, a set of five two-qubit gates suffices to implement the DQS:
\begin{eqnarray}
g_1 &=& \exp\left(\upi\frac{\pi}{4} Y_3X_4\right),\nonumber\\
g_2 &=& \exp\left(\upi\frac{\pi}{4} Z_3Y_2\right),\nonumber\\
g_3 &=& \exp\left(\upi\varphi X_1Z_2\right),\nonumber\\
g_4 &=& \exp\left(-\upi\frac{\pi}{4} Z_3Y_2\right)=g_2^{-1} ,\nonumber\\
g_5 &=&\exp\left(-\upi\frac{\pi}{4} Y_3X_4\right) =g_1^{-1}.
\end{eqnarray}
Here, $X_i, Y_i$ and $Z_i$ are the Pauli operators acting on the $i$th qubit. Observe that
\begin{equation}
g_5\cdot g_4\cdot g_3 \cdot g_2 \cdot g_1 = \exp(i\varphi X_1X_2X_3X_4)= \upe^{-\upi\HStar T},
\end{equation}
with $T \equiv 2\varphi / \omega$ being the simulation cycle period. The gate sequence $g_5\ldots g_2g_1\equiv \USsim(T,0)$ equals to $\UStar(T,0)$, and repeated sequences, implemented with cycle time $T$, achieve exact simulation of the target Hamiltonian $\HStar$. The five gates are applied one after another in sequence, each as an instantaneous pulse, separated equally in time and taking a total time $\tau_g=RT$ ($R\equiv R_5$ for our 5-gate DQS) to complete. Such a set of gates may, in practice, also be difficult to implement, but here, we are only concerned with using it as a platform for studying the fidelity of multi-gate simulation in the presence of a bath.

We suppose that the system (target or simulator) is coupled to an oscillator bath (as in Sec.~\ref{sec:0_temp_ohmic_bath}) with an Ohmic spectral density. Each qubit is assumed to interact with an independent oscillator bath, as would be the situation if the distance between pairs of qubits is large compared to the correlation length of the bath, and the bath degrees of freedom coupled to different qubits do not interact. Since $\HStar$ protects against $Z$ errors in the system, we take $A_k=Z_k$, so that
\begin{eqnarray}\label{eq:coupling}
\HSB&=&\sum_{k=1}^4 Z_k\otimes B_k,\\
\textrm{with } B_k&=&\sum_m \left(g_{k,m} b_{k,m}^\dagger + g_{k,m}^* b_{k,m}\right),\nonumber
\end{eqnarray}
where $b_{k,m}$ $(b_{k,m}^\dagger)$ is the annihilation (creation) operator for the $m$th mode of the oscillator bath that interacts with qubit $k$. The Ohmic spectral density is [see Eq.~\eqref{eq:OhmicSpec}]
\begin{equation}\label{eq:coupling2}
\cJ_{k\ell}(\nu)= \delta_{k\ell}\,\eta\, \nu \,\upe^{-\nu/\nu_c},
\end{equation}
where the Kronecker delta $\delta_{k\ell}$ encapsulates the independent bath assumption. Correspondingly, the bath correlation function satisfies $f_{k\ell}(s)=\delta_{k\ell}f_k(s)$, where $f_k(s)\equiv f_{kk}(s)$.

\begin{figure*}
\includegraphics[width=\textwidth]{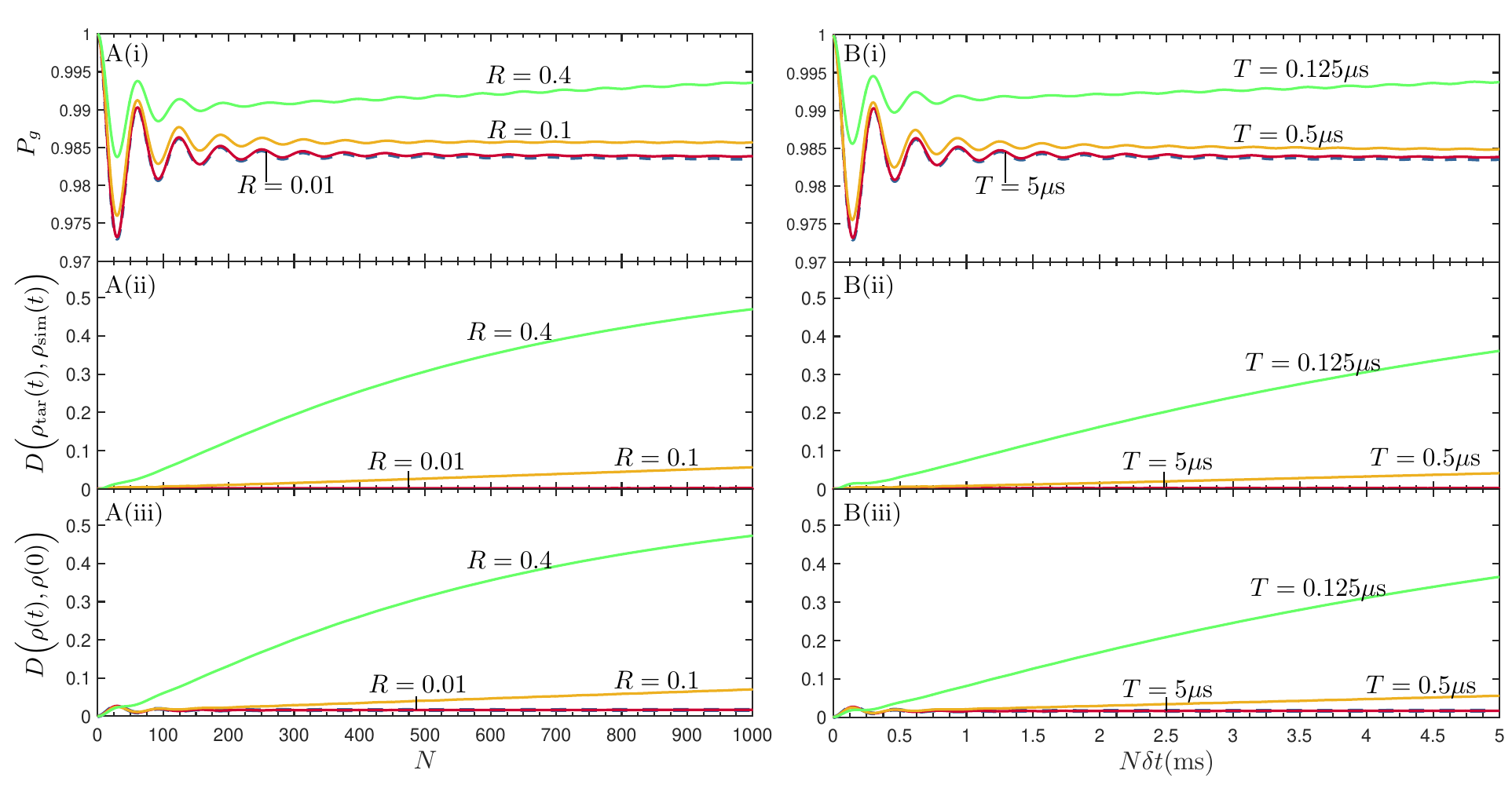}
\caption{\label{fig4}
Dynamics of the toric-code vertex target and simulator, in the regime where $x_c\ll \epsilon$, or, equivalently, where $\omega\tauB\ll1$. The left column [plots marked (A)] gives the situation where all parameters but $R=\tau_g/T$ are fixed (to be regarded as varying $\tau_g$ for fixed $T$; see main text). The right column [plots marked (B)] gives the case where all parameters but the simulation cycle time $T$ are fixed. Plots marked (i) give the ground-space population; those marked (ii) give the trace distances between the target and simulator states; those marked (iii) are the trace distances between the time $t$ state $\rho(t)$ and the initial state, for the target and the simulator. Parameters for plots A: $\epsilon=0.1, x_c=\epsilon/5=0.02,\widetilde\eta=0.02$, and $\widetilde\beta=40$. Parameters for plots B: $\omega=20$kHz, $\nu_c=4$kHz, $\beta=0.2$ms, $\eta=0.02$, and $\tau_g=50$ns. The blue dashed line is for the target. In plot A(ii), the $R=0.01$ line essentially lie on the horizontal axis, for the plotted vertical scale; in plot B(ii), the $T=5\mu$s line is also nearly on the horizontal axis.}
\end{figure*}

We calculate the density matrix for both the target and the simulator using the second-order (in $\alpha \HSB$) time-convolutionless master equation (TCL-2) for the open-system dynamics \cite{Breuer:2002:OxfordUniversityPress},
\begin{eqnarray}\label{eq:TCL-2_simulation}
&&\frac{\upd}{\upd t}\rho_S(t)\\
&=&\sum_k \int_{0}^t \upd s{\left[f_k(t-s)-f_k(s-t)\right]} A^{(\mu)}_k(s){\left[\rhoS(t),A^{(\mu)}_k(t)\right]}\nonumber
\end{eqnarray}
for $\mu=\Tar,\Sim$. TCL-2 is a non-Markovian master equation, valid in the weak-coupling limit (i.e., small $\alpha$). 
We solve Eq.~\eqref{eq:TCL-2_simulation} using a fourth-order Runge-Kutta method \cite{numerical_recipe_press} for the density matrices $\rhoS(t)$ of the target and the DQS, for the initial $(t=0)$ GHZ-type state $\frac{1}{\sqrt{2}}{\left(|0000\rangle+|1111\rangle\right)}$ in the ground space (code space). Here, $|0\rangle$ and $|1\rangle$ are the eigenstates of $Z$, with eigenvalues $+1$ and $-1$, respectively. The numerical results are observed to depend only very weakly on which state from the ground space is chosen as the initial state, so the above state suffices to illustrate the point. Being an approximate equation, TCL-2 does not in general guarantee a positive, unit-trace $\rhoS(t)$, especially for long-time evolution, but we see no such defects within the time period of our numerical calculations.

\subsubsection{$x_c=\nu_cT =T/\tauB\ll \epsilon=\omega T \ll 1$}\label{sec:nMsimulation}

We first study the regime (with $\bar x=0$) where $x_c\ll \epsilon$, i.e., $\omega\tauB\gg 1$ so that the target system timescale ($\sim1/\omega$) is much smaller than that of the bath. Fig.~\ref{fig4} shows the stroboscopic dynamics of the target and the DQS. We vary the two parameters within the control of the simulator design: the total sequence time $\tau_g$ (Fig.~\ref{fig4}A) and the simulation cycle time $T$ (Fig.~\ref{fig4}B).

In Fig.~\ref{fig4}A, we set  $\epsilon= 0.1$ and $x_c=\epsilon/5=0.02$. For weak coupling between the system and the bath, we set $\widetilde\eta=0.02$ ($=\eta$ in the Ohmic case where $w=1$). We consider a nonzero temperature to observe the effects of thermal noise by putting $\widetilde\beta\equiv\beta/T=40$. Note that $T$ here is the simulation cycle time, \emph{not} the temperature; $k$, the Boltzmann constant is set to 1 so that the inverse temperature $\beta$ has dimensions of frequency (with $\hbar$ also set to 1). Observe that the ratio of the temperature to the target system energy scale, given by $1/(\omega\beta) =1/(\epsilon\widetilde\beta)=0.25$, is small compared to 1, so we are in a somewhat low-temperature regime. The parameter $R=\tau_g/T$ is varied over $0.01, 0.1$ and $0.4$. These fixed or varying dimensionless parameter values should be understood as follows: The physical parameters $\omega, \nu_c,\eta$ and $\beta$ are determined by the system and the bath under consideration, so fixing the values of $\epsilon,x_c, \widetilde\eta$ and $\widetilde\beta$ means that $T$ is fixed; consequently, varying $R$ is equivalent to varying $\tau_g$. 

At zero-temperature, $\epsilon\widetilde f(0)=\epsilon\widetilde \eta x_c^2=8\times 10^{-7}$, so that, naively, Eq.~\eqref{cond12} gives $N\lesssim 1000$ for small simulation error; we might expect the low-temperature case to be similar. However, note that $N\epsilon\sim 100$ and $N x_c\sim 20$ for $N\sim 1000$, neither of which are small. Thus, we are not in the regime of our analyses of Secs.~\ref{sec:1gateDQS} and \ref{sec:MgateDQS}, where we assumed also that $N\epmax$ and $Nx_c\ll 1$. In fact, we leave that regime once $N\gtrsim 10$, very early on in the numerical simulation below.

Figure~\ref{fig4}A plots the dynamics of the target and the DQS in steps of $T$. One observes that
the ground-space population [see Fig.~\ref{fig4}A(i)] for the target system oscillates early on, but reaches a steady level close to 1 in the long-time regime. This behaviour is indicative of typical non-Markovian dynamics. That the ground-space population remains always close to 1 demonstrates the ability of the system to suppress the leakage effects of the environmental coupling out of the ground space: The long-time ratio between the transition rate into the code space and the rate out of the code space is $\upe^{\beta\omega}=\upe^4\simeq 54$.

For small values of $R$ ($R=0.01$ and $0.1$), the state of the simulator remains close to that of the target system [see Fig.~\ref{fig4}A(ii)], and the behavior in terms of the ground-space population is similar. When $R$ gets larger, one starts to see deviation of the DQS from the target, as is clearly visible from the $R=0.4$ case in Fig.~\ref{fig4}A. One thus has better simulation for smaller $R$, and this extends our conclusions of Sec.~\ref{sec:MgateDQS} to beyond the analytically accessible regimes. 

An interesting feature noticeable in Fig.~\ref{fig4}A(i) is that larger $R$ actually gives larger ground-space population. As there is no reason to suspect that the numerical errors are larger for larger $R$, the plots suggest that, as far as keeping the system in the code space is concerned, the larger-$R$ simulators seem to perform better. However, Fig.~\ref{fig4}A(iii) indicates that larger $R$ values result in greater in-code-space operations, i.e., logical errors, on the system state, which are harmful as far as preservation of the logical information is concerned. Such operations are neither detectable, nor correctable, by the code. Hence, if the intention is solely to keep the population in the ground space, larger $R$ works better, but not if one also wants to preserve the particular state of the system.

In Fig.~\ref{fig4}B, we are varying $T$ itself, which we have been using as the time unit for our dimensionless variables. Thus, the results are reported for specific fixed values of physical parameters $\omega,\nu_c,\eta$ and $\beta$, as well as a given $\tau_g$ value, while $T$ is varied. We plot the dynamics in timesteps of $\delta t=5\mu$s (the largest $T$ value), and vary $T$ from 125 ns to 5$\mu$s. In Sec.~\ref{sec:MgateDQS}, we saw that the simulation error for the $M$-gate simulator $\sS_M$ had two contributions, one that goes as $1/T$ [from $\Err{N}(\sS_1,\sS_M)$], the other as $T$ [from $\Err{N}(\Tar,\sS_1)$]. There, we could not come to a definitive conclusion about the overall behavior of the simulation error as a $T$ changes, as the relative weights of the two contributions depend on the problem. Here, for our toric-code vertex example, Fig.~\ref{fig4}B shows that the overall simulation error \emph{decreases} as $T$ increases, indicating that the $1/T$ term wins. This is contrary to conventional wisdom where one expects more rapid repetition of the simulator sequence to give better performance. Here, to better mimic the dynamics of the target, one should instead wait for a period of time between consecutive gate sequences, so that $T$ is larger, at least as long as $T$ remains small enough such that $\epsilon=\omega T\ll1 $ for good simulation [Condition~\eqref{cond:shortT}].

\begin{figure}
\centering\includegraphics[width=\columnwidth]{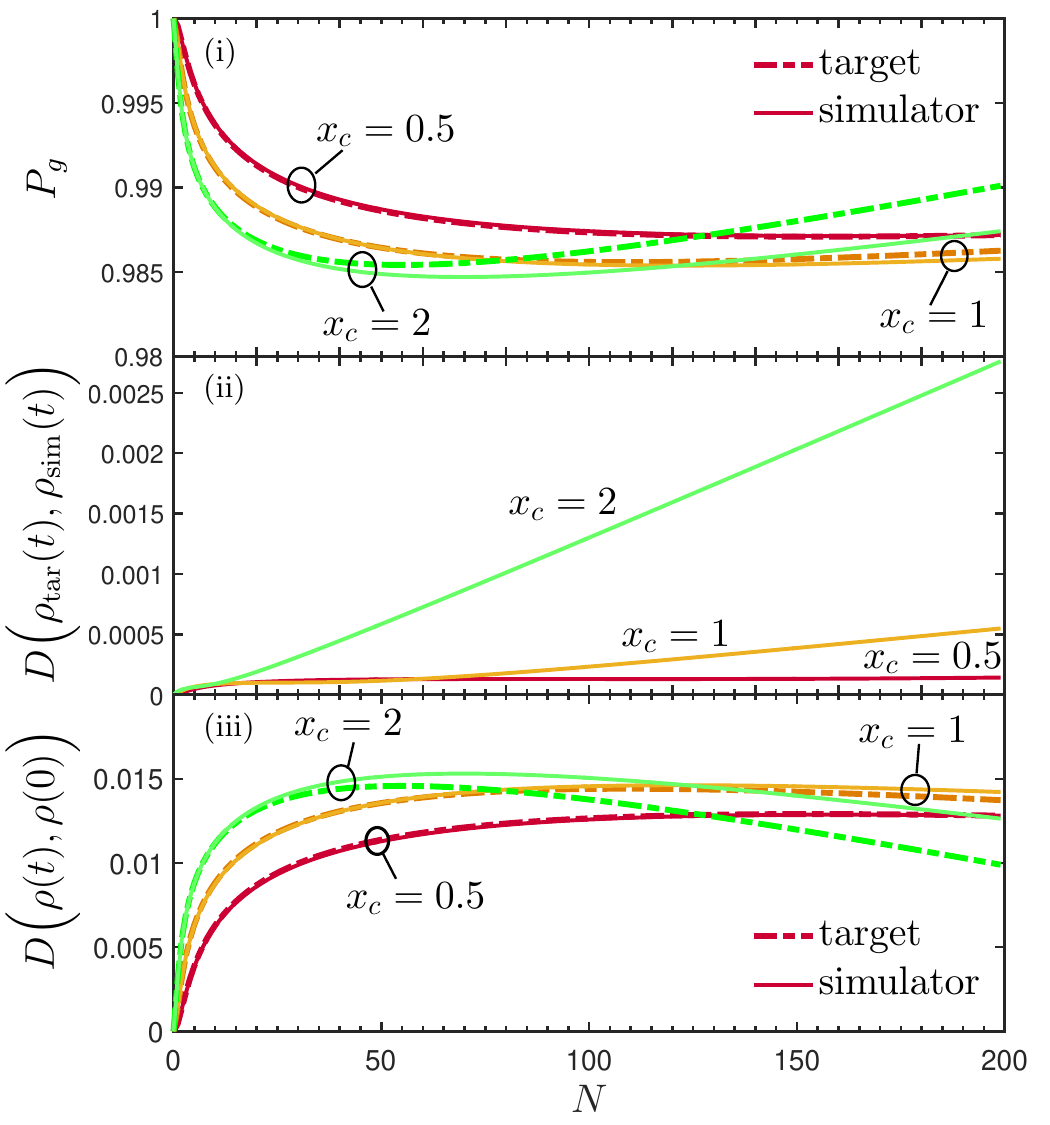}
\caption{\label{fig:Long}The dynamics of the target and the DQS, with varying $x_c=\nu_cT=T/\tauB\sim 1$ values. The other parameters are kept fixed: $\epsilon = 0.005, \widetilde\eta=5\times 10^{-4}$, $\widetilde\beta=2000$, and $R=0.01$. The plots are labeled by $x_c/\epsilon=100,200$ and $400$, corresponding to $x_c=0.5, 1$ and $2$, respectively.
}
\end{figure}

\begin{figure*}
\includegraphics[width=\textwidth]{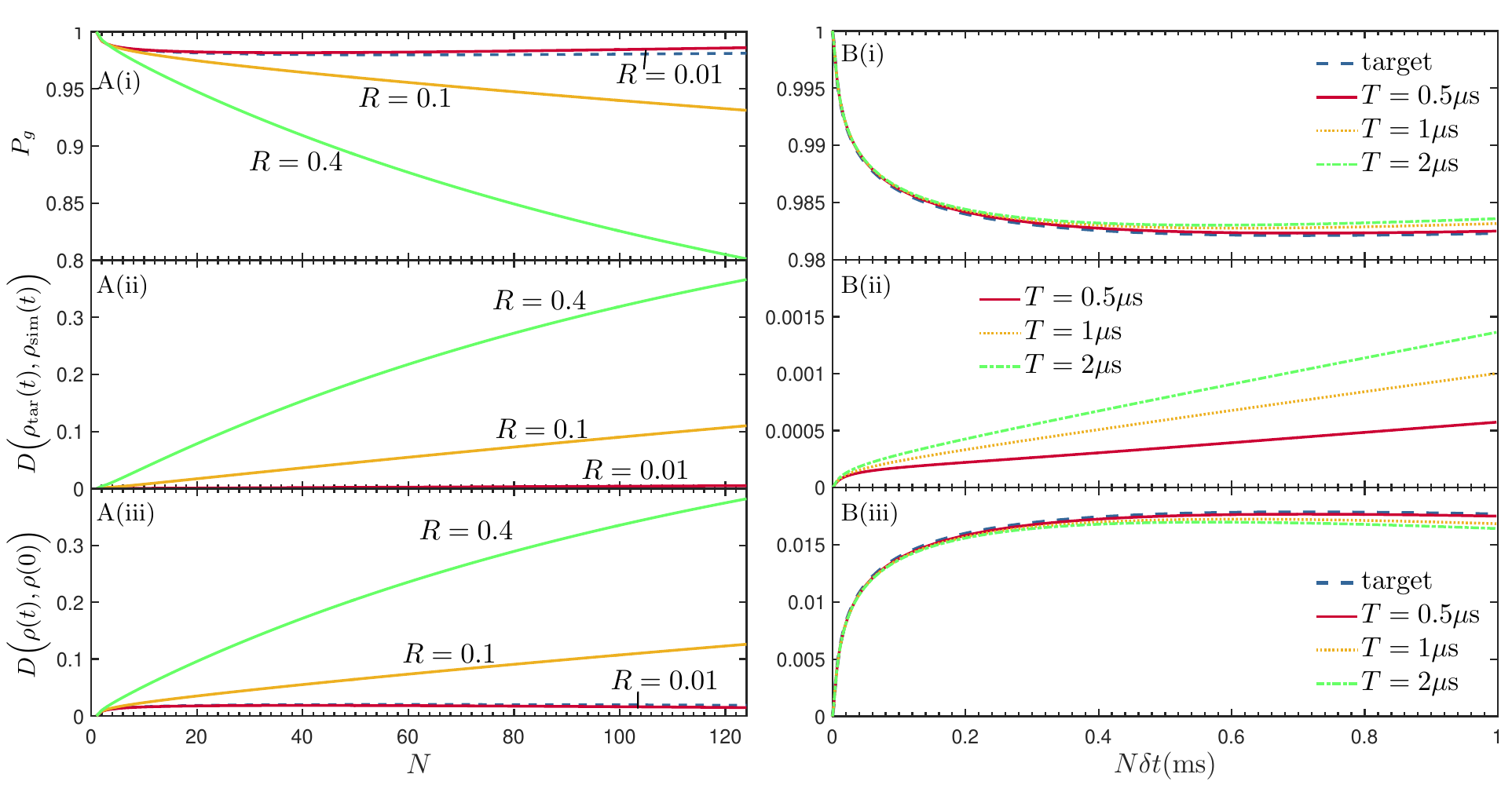}
\caption{\label{fig6}
Dynamics of the toric-code vertex target and the simulator, in the regime where $x_c\gg \epsilon$, or, equivalently, where $\omega\tauB\gg1$. The left column [plots marked (A)] gives the situation where all parameters but $R=\tau_g/T$ are fixed. The right column [plots marked (B)] gives the case where all parameters but the simulation cycle time $T$ and $\tau_g$ (with $R=\tau_g/T$ held constant) are fixed. Plots marked (i) give the ground-space population; those marked (ii) are the trace distance between the target and simulator states; those marked (iii) are the trace distance between the time $t$ state $\rho(t)$ and the initial state, for the target and the simulator. Parameters for plots A: $\epsilon = 4\times 10^{-4}$, $x_c=8\gg \epsilon$, $\widetilde\eta=5\times 10^{-4}$, and $\widetilde\beta=2500$. Parameters for plots B: $R$ is fixed to be 0.01, $\omega=1$kHz, $\nu_c=400$kHz, $\beta=1$ms, $\eta=5\times 10^{-4}$, and $\delta t=2\mu$s.
}
\end{figure*}

\subsubsection{$ \epsilon=\omega T  \ll x_c=\nu_cT=T/\tauB$}

Let us examine a different parameter regime, where $\epsilon \ll x_c$, or, equivalently, $\omega\tauB\ll 1$, i.e., the target system timescale is much larger than that of the bath. This is often referred to as the Markovian regime in the weak-coupling limit. First, we focus on the case where $x_c\sim 1$, where the analytical estimation was difficult and lacking. Figure \ref{fig:Long} shows the stroboscopic dynamics for three different $x_c$ values: $x_c=0.5,1,$ and 2. All other parameters are kept fixed: $\epsilon=0.005$, $\widetilde\eta=5\times 10^{-4}$, and $\widetilde{\beta}=2000$. Note that $1/(\epsilon\widetilde\beta)=0.1$, so we are in the low-temperature range. In all three cases, the simulation errors are small: Observe that the trace distance between the target and simulator states are no larger than $\sim10^{-3}$ in the time shown. The error is noticeably larger for larger $x_c$, and given the growing trend, one expects the simulation error to eventually become significant, but only at long times (large $N$).

Next, one can study the effects of changing $\tau_g$ and $T$. Figure \ref{fig6} shows the dynamics of the target and the simulator in the regime of $\epsilon\ll x_c$, i.e., $\omega\tauB\ll1$, for different $\tau_g$ (Fig.~\ref{fig6}A) and $T$ (Fig.~\ref{fig6}B) values. 
The plots of Fig.~\ref{fig6}B show what one might typically expect (unlike the situation in Fig.~4B), that the simulation error increases as $T$ increases.

\subsection{The five-qubit code}
The toric-code vertex model of the previous subsection can only suppress $Z$ errors in one qubit (or $X$ errors if one uses $Z$ operators in $\HStar$). Here, we consider the five-qubit code~\cite{Laflamme:1996:198}, the smallest-sized code capable of correcting an arbitrary error on any one of the qubits. The target Hamiltonian in this case is
\begin{equation}\label{eq_5}
\HStar=-\gamma\sum_{j=1}^4 S_{j}
\end{equation}
where $S_j$ are the stabilizer generators of the five-qubit code,
\begin{eqnarray}
S_1 &= & X_1Z_2 Z_3 X_4,\nonumber\\
S_2 &= & X_2Z_3 Z_4 X_5,\nonumber\\
S_3 &= & X_1X_3 Z_4 Z_5, \nonumber\\
S_4 &= & Z_1X_2 X_4 Z_5.
\end{eqnarray}
The two-dimensional ground space of $\HStar$ forms the qubit codespace, with the logical $X$ and $Z$ operators chosen to be $X_1X_2X_3X_4X_5$ and $Z_1Z_2Z_3Z_4Z_5$, respectively. Any single-qubit error will cause a transition from the ground space to the higher-energy excited space, and correspondingly, such an error will be energetically unfavorable and suppressed in this model.

\begin{table*}[!htp]
\label{tab:gate_sequence}
\begin{center}
\begin{tabular}{c| c |c |c }
  \hline
  \hline
  $S_1=X_1Z_2Z_3X_4$ & $S_2=X_2Z_3Z_4X_5$ & $S_3=X_1X_3Z_4Z_5$ & $S_4=Z_1X_2X_4Z_5$ \\
  \hline
  $\quad g_1 =\exp\left(\upi \frac{\pi}{4}X_1X_2\right)\quad$ & $\quad g_6=\exp\left(\upi \frac{\pi}{4}Z_3Y_5\right)\quad$& $\quad g_{11}=\exp\left(\upi \frac{\pi}{4}X_1Y_3\right)\quad$ & $\quad g_{16}=\exp\left(\upi \frac{\pi}{4}X_4X_5\right)\quad$ \\[2pt]
  $g_2 =\exp\left(\upi \frac{\pi}{4}Y_2X_3\right)$ & $g_7 =\exp\left(\upi \frac{\pi}{4}Z_4Z_5\right)$ & $g_{12}=\exp\left(\upi \frac{\pi}{4}Z_3X_4\right)$ & $g_{17}=\exp\left(\upi \frac{\pi}{4}Y_2Y_5\right)$ \\[2pt]
  $g_3 =\exp\left(\upi \varphi Y_3 X_4\right)$ & $g_8=\exp\left(\upi \varphi X_2 Y_5\right)$ & $g_{13}=\exp\left(\upi \varphi Y_4 Z_5\right)$ & $g_{18}=\exp\left(\upi \varphi Z_1 Z_2\right)$ \\[2pt]
  $g_4 =g_2^{-1}$& $g_9=g_7^{-1}$ & $g_{14}=g_{12}^{-1}$ & $g_{19}=g_{17}^{-1}$ \\[2pt]
  $g_5 =g_1^{-1}$ & $g_{10}=g_6^{-1}$ & $g_{15}=g_{11}^{-1}$ & $g_{20}=g_{16}^{-1}$ \\
  \hline
  \hline
\end{tabular}
\caption{\label{tab:5_qubit_gates} The gate sequence for simulation of the five-qubit code, with $\upe^{-\upi H_{\mathrm{tar}}T}=g_{20}g_{19}\,\cdots\, g_2 g_1$. Here, $\varphi=\gamma T$.}
\end{center}
\end{table*}

As in the case of the toric-code vertex, one can build an exact DQS of this $\HStar$ from a set of two-qubit gates. The specific set of gates---twenty gates in all---we use is given in Table~\ref{tab:5_qubit_gates}.
Note that these gates are not chosen with any particular implementation in mind, and are used here solely for the purpose of examining the dependence of the simulation error on various physical parameters.

We again study this five-qubit code situation using numerical solution of the TCL-2 master equation, for the same noise as for our toric-code vertex example, i.e., an Ohmic oscillator-bath noise described by Eqs.~\eqref{eq:coupling} and \eqref{eq:coupling2}. The initial state is taken to be the logical 0 state of the code,
\begin{equation}\label{eq:phi_0_5qubit}
|\bar 0\rangle= \frac{1}{\sqrt{8}}(\id + S_1)\, (\id + S_2)\,  (\id+S_3)\,(\id+S_4)\,|00000\rangle.
\end{equation}
As in the previous example, we see little variation numerically for different initial states; this choice hence suffices for illustration. Note that the five-qubit code is capable of protecting against arbitrary single-qubit errors, even though the noise coming from $\HSB$ has only Pauli $Z$ operators.

\begin{figure*}
\includegraphics[width=\textwidth]{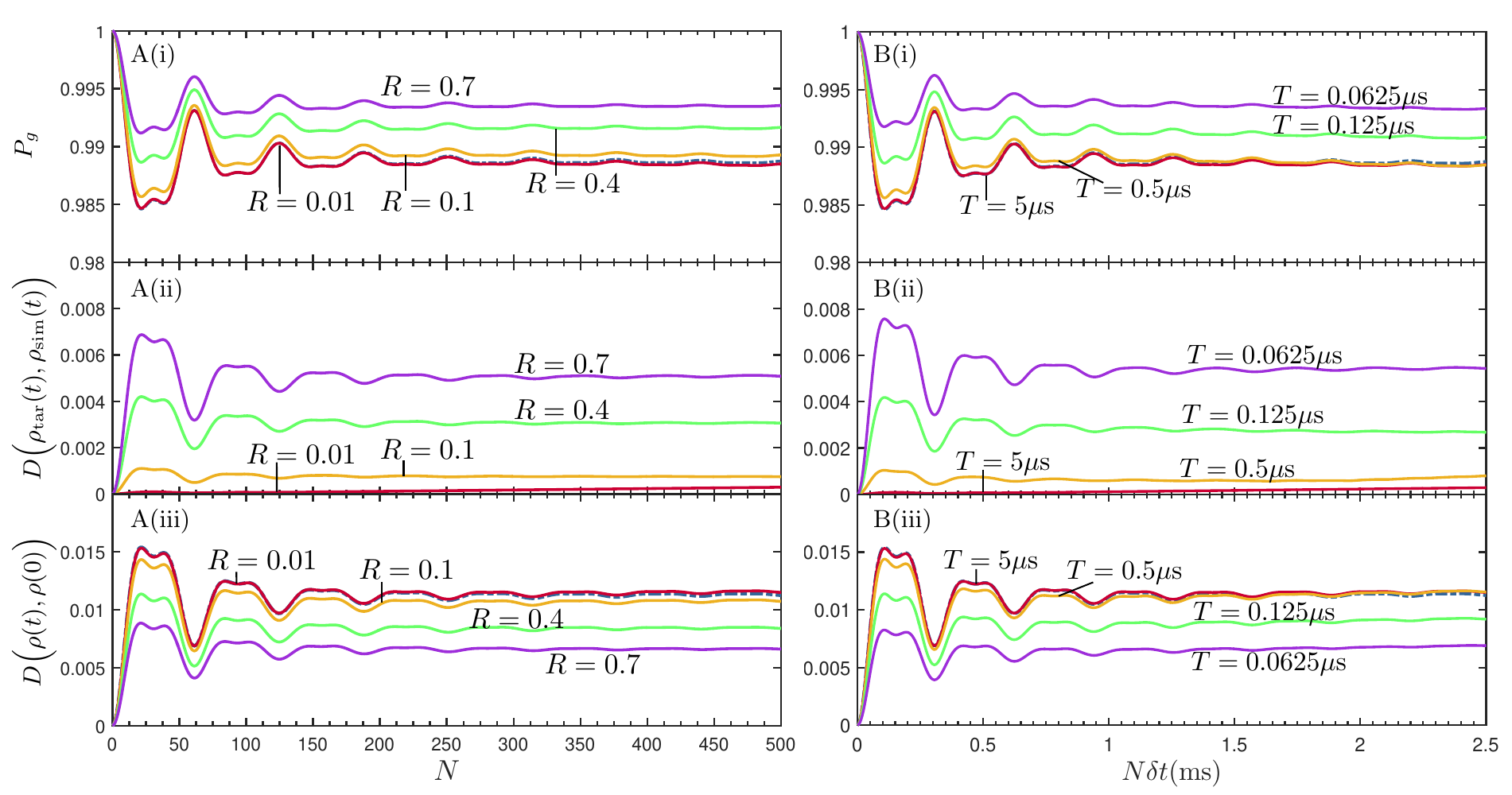}
\caption{\label{fig7}
Dynamics of the five-qubit code target and the simulator, in the regime where $x_c\ll \epsilon$, or, equivalently, where $\omega\tauB\ll1$. (This is similar to Fig.~\ref{fig4}, but for the five-qubit code, rather than the toric-code vertex.) The left column [plots marked (A)] gives the situation where all parameters but $R=\tau_g/T$ are fixed; the right column [plots marked (B)] gives the case where all parameters but the simulation cycle time $T$ are fixed. Plots marked (i) give the ground-space population; those marked (ii) are the trace distance between the target and simulator states; those marked (iii) are the trace distance between the time $t$ state $\rho(t)$ and the initial state, for the target and the simulator. Parameters for plots A: $\epsilon=0.1$, $x_c=0.2$, $\widetilde\eta=0.02$ and $\widetilde\beta=40$; $R=\tau_g/T$ is varied over the values $0.01, 0.1, 0.4$ and $0.7$ (to be regarded as varying $\tau_g$ for fixed $T$). Parameters for plots B: $\omega=20$kHz, $\nu_c=4$kHz, $\beta=0.2$ms, $\eta=0.02$, $\tau_g=50$ns, and $\delta t=5\mu$s.
}
\end{figure*}

Rather than examine a variety of situations as we did for the toric-code vertex example, we focus here on the regime where $x_c\ll \epsilon \ll 1$ and on the effect of different values of $R$ for DQSs. As we will see below, the behavior of the five-qubit code is somewhat different from that of the toric-code vertex with similar parameters. As in Fig.~\ref{fig4}A, we set $\epsilon=0.1$, $x_c=0.02$, $\widetilde{\eta}=0.02$ and $\widetilde{\beta}=40$. Fig.~\ref{fig7}A shows the stroboscopic dynamics of the target and the simulator, with points plotted every time-step $T$. As $R$ increases (i.e., $\tau_g$ increases with $T$ fixed), the simulation error, as captured by the trace distance between the target and simulator states [see Fig.~\ref{fig7}A(ii)], increases, much like what was observed in Fig.~\ref{fig4}A, reaffirming our earlier conclusions. 

What is unexpected, and dissimilar from the toric-code vertex example, is that the error does not grow with time but stabilizes to some value at long times. What is even more surprising are the plots of Fig.~\ref{fig7}A(iii): The deviation of the simulator state from its initial state is \emph{smaller} for larger $R$, even though the larger-$R$ DQS does a poorer job of imitating the target.
We do not know the source of this effect and it may deserve further exploration, but it is beyond the scope of our current discussion.

For completeness, we also present the case of varying $T$ with all other physical parameters held fixed; see Fig.~\ref{fig7}B. As for the toric-code vertex, Fig.~\ref{fig7}B provides evidence that a larger $T$ leads to smaller simulation errors.

\section{Summary and discussion}\label{sec:Conc}
We compared, analytically and numerically, the stroboscopic dynamics of a DQS and its target system in the presence of a bath. It is clear that the simulator and target dynamics are similar under a combination of conditions, as summarized in Table \ref{tab:condition} for limiting physical regimes. The common belief that $T$ should always be short for good simuation is neither sufficient---for example, one also needs $f(0)$ to also be small for $\sS_1$ in regimes I and II---nor necessary---our example of the toric-code vertex demonstrates a situation where larger $T$ incurs a smaller error.
Under the conditions where the DQS and target are similar, the simulation pulse sequences successfully suppress the errors in the system, providing effective robustness to noise from the bath.

In our work, we have assumed that the applied gates for the DQS are instantaneous. This is a good approximation for many experimental architectures currently in play. 
It is interesting to note that the periodic sequences of fast gate pulses for DQS are reminiscent of the technique of dynamical decoupling (DD)~\cite{Viola:1998:2733,Viola:2000:3520,Khodjasteh:2005:180501,Ng:2011:012305}. DD aims for a vanishing effective Hamiltonian on the system (to the order of the DD sequence). This can been viewed as a special case of DQS, with a zero target Hamiltonian. In DD, the usual requirements are that the pulses are fast, and the cycle time is short. One wonders if there is also a situation in which a larger cycle time gives better results, as has been seen to be possible in our work. 

Going forward, one can ask for a more careful, but no doubt more complicated, analysis where the simulator gates pulses are not instantaneous. This introduces one more timescale into the problem, which should enter the simulation error. The design error, set to be zero in our work so as to focus on the impact of the environment, is also realistically nonzero, and one could take it into account in the calculation. This includes the gate-pulse errors, which can be regarded as a special type of design error.

In addition, one could also go away from the static target Hamiltonian we have restricted ourselves to here, and look into slowly varying target Hamiltonians. Again this adds one more timescale to the problem, but it is a worthy subject for further studies, as it goes towards schemes of Hamiltonian-based quantum computation~\cite{JordanFarhiShorPhysRevA.74.052322,Bacon-AGTPhysRevLett.103.120504,
bacon2010adiabaticcluster,Yi-Cong_PhysRevA.89.032317,
zheng2015fault_hqc_surface,cesare2015adiabatic}.

Another potential application of our results is in the preparation of the system in the equilibrium state of a complicated many-body target Hamiltonian. The existing quantum algorithm of Gibbs preparation relies on quantum phase estimation~\cite{poulin2009sampling_gibbs}. Instead, one could imagine coupling the DQS to a thermal bath at temperature $1/\beta$, that of the Gibbs state we want to prepare. If one can fulfil the condition to ``cheat the bath'' into seeing the target Hamiltonian as the effective Hamiltonian for the system, then the bath will thermalize the system to the equilibrium state of the target many-body Hamiltonian (provided it is in the Markovian regime, and the quantum semi-group has the ergodic property~\cite{Breuer:2002:OxfordUniversityPress,rivas2012open_book}). This method may be considered as a noise-assisted state preparation. 
Our work provides the conditions under which this approach would be successful.

\acknowledgments
This work is supported by the National Research Foundation of Singapore and Yale-NUS College (through grant number IG14-LR001 and a startup grant).

\bigskip
\bigskip

\appendix

\section{Derivation of $\cE_{N,I}^{(\Tar)}-\cE_{N,I}^{(\Sim)}$}\label{app:EDiff}
We begin with the approximation of $U^{(\mu)}_I(t)$ ($\mu=\Tar,\Sim$) to second order in $\alpha$ [Eq.~\eqref{eq:UI} in the main text]:
\begin{eqnarray}
U_I^{(\mu)}(t)&\simeq& \id -\upi\alpha\int_0^t \upd s \HSB^{(\mu)}(s)\\
&&\quad -\alpha^2\int_0^t \upd s\int_0^s \upd s' \HSB^{(\mu)}(s)\HSB^{(\mu)}(s').\nonumber
\end{eqnarray}
We first gather the relevant relations from the main text: $\HSB=\sum_kA_k\otimes B_k$, where $A_k$ acts on the system, and $B_k$ on the bath; $A_k^{(\mu)}(t)\equiv U_\mu(t)^\dagger A_k U_\mu(t)$, and $B_k(t)\equiv U_\mathrm{B}(t)^\dagger B_k U_\mathrm{B}(t)$, the interaction-picture operators; $\langle B \rangle\equiv \Tr(B\rhoB)$, for any bath-only operator $B$, and $\rhoB$ the initial bath state; $\rhoB$ is a stationary state of $\HB$, i.e., $[\HB,\rhoB]=0$, and $\langle B_k(t)\rangle=0\forall k,t$; the two-point bath correlation function $f_{k\ell}(t,s)\equiv\langle B_k(t)\,B_\ell(s)\rangle$, for which $f_{k\ell}(t,s)^*=f_{\ell k}(s,t)$;
stationarity means that $f_{k\ell}(t,s)=f_{k\ell}(t-s,0)\equiv f_{k\ell}(t-s)$, and thus $f_{k\ell}(t)^*=f_{\ell k}(-t)$.

Using these, straightforward algebra yields
\begin{widetext}
\begin{eqnarray}
\cE_{N,I}^{(\mu)}(\,\cdot\,)&=&\Tr_\mathrm{B}{\left(U_I^{(\mu)}(NT)(\,\cdot\,\otimes\rhoB)U_I^{(\mu)}(NT)^\dagger\right)}\nonumber\\
&=&\id+\alpha^2\sum_{k\ell}{\left\{\int_0^{NT}\upd t\int_0^{NT}\upd s\,f_{k\ell}(t-s)A_\ell^{(\mu)}(s)(\,\cdot\,)A_k^{(\mu)}(t)\right.}\nonumber\\
&&\quad{\left.-\int_0^{NT}\upd t\int_0^{t}\upd s\,f_{k\ell}(t-s)A_k^{(\mu)}(t)A_\ell^{(\mu)}(s)(\,\cdot\,)-\int_0^{NT}\upd t\int_0^{t}\upd s\,(\,\cdot\,)A_\ell^{(\mu)}(s)A_k^{(\mu)}(t)\,f_{k\ell}(t-s)^*\right\}}.\label{eq:ENImu}
\end{eqnarray}
$\cE^{(\Tar)}_{N,I}-\cE^{(\Sim)}_{N,I}$ is a sum of three maps, each of order $\alpha^2$,
\begin{equation}\label{eq:channel_difference}
\cE^{(\Tar)}_{N,I}-\cE^{(\Sim)}_{N,I}=\alpha^2(\Delta_1+\Delta_2+\Delta_3),
\end{equation}
where $\Delta_i(\cdot)$ is the difference between $\mu=\Tar$ and $\mu=\Sim$ of the $i$th non-identity terms of Eq.~\eqref{eq:ENImu}.
For example, $\Delta_2(\cdot)$ is given by
\begin{eqnarray}
&&\Delta_2(\,\cdot\,)\\
&=&-\sum_{k\ell}\int_0^{NT}\upd t\int_0^{t}\upd s\,f_{k\ell}(t-s){\left[A_k^{(\Tar)}(t)A_\ell^{(\Tar)}(s)-A_k^{(\Sim)}(t)A_\ell^{(\Sim)}(s)\right]}(\,\cdot\,)\nonumber\\
&=&-\sum_{k\ell}\int_0^{NT}\upd t\int_0^{t}\upd s\,f_{k\ell}(t-s){\left\{A_k^{(\Tar)}(t){\left[A_\ell^{(\Tar)}(s)-A_\ell^{(\Sim)}(s)\right]}+{\left[A_k^{(\Tar)}(t)-A_k^{(\Sim)}(t)\right]}A_\ell^{(\Sim)}(s)\right\}}(\,\cdot\,)\nonumber\\
&=&-\sum_{k\ell}{\left\{\int_0^{NT}\!\!\!\!\!\upd t\!\int_0^{t}\!\!\!\upd s\,f_{k\ell}(t\!-\!s)A_k^{(\Tar)}(t){\left[A_\ell^{(\Tar)}(s)\!-\!A_\ell^{(\Sim)}(s)\right]}
+\!\int_0^{NT}\!\!\!\!\!\upd s\!\int_0^{s}\!\!\!\upd t\,f_{\ell k}(s\!-\!t){\left[A_\ell^{(\Tar)}(s)\!-\!A_\ell^{(\Sim)}(s)\right]}A_k^{(\Sim)}(t)\right\}}(\,\cdot\,)\nonumber\\
&=&-\sum_{k\ell}{\left\{\!\int_0^{NT}\!\!\!\!\!\upd t\!\int_0^{t}\!\!\!\upd s\,f_{k\ell}(t\!-\!s)A_k^{(\Tar)}(t){\left[A_\ell^{(\Tar)}(s)\!-\!A_\ell^{(\Sim)}(s)\right]}
+\!\int_0^{NT}\!\!\!\!\!\upd t\!\int_t^{NT}\!\!\!\!\!\upd s\,f_{k\ell}(t\!-\!s)^*{\left[A_\ell^{(\Tar)}(s)\!-\!A_\ell^{(\Sim)}(s)\right]}A_k^{(\Sim)}(t)\!\right\}}(\,\cdot\,).\nonumber
\end{eqnarray}
In the last line, we have used the fact that $\int_0^{NT}\upd s\int_0^s\upd t \,F(t,s)=\int_0^{NT}\upd t\int_t^{NT}\upd s\, F(t,s)$ for any function $F$, and that $f_{\ell k}(x)=f_{k\ell}(-x)^*$.
We switch to dimensionless quantities, with integration variables $a\equiv s/T$ and $b\equiv t/T$. Then, one can write $\Delta_2(\cdot)$ as
\begin{eqnarray}
\Delta_2(\,\cdot\,)&=&-\sum_{k\ell}\int_0^{N}\upd b\,\widetilde A_k^{(\Tar)}(b)\,{\left\{\int_0^{b}\upd a\,\widetilde f_{k\ell}(b-a)\,{\left[\widetilde A_\ell^{(\Tar)}(a)-\widetilde A_\ell^{(\Sim)}(a)\right]}\right\}}(\,\cdot\,)\nonumber\\
&&\quad-\sum_{k\ell}\int_0^{N}\upd b\,{\left\{\int_b^{N}\upd a\,\widetilde f_{k\ell}(b-a)^*\,{\left[\widetilde A_\ell^{(\Tar)}(a)-\widetilde A_\ell^{(\Sim)}(a)\right]}\right\}}\,\widetilde A_k^{(\Sim)}(b)(\,\cdot\,)\,.
\end{eqnarray}
Defining, as in the main text,
\begin{eqnarray}
\Lambda_{k\ell}(b)&\equiv& \int_0^{b}\upd a\,\widetilde f_{k\ell}(b-a)\,{\left[\widetilde A_\ell^{(\Tar)}(a)-\widetilde A_\ell^{(\Sim)}(a)\right]},\nonumber\\
\overline\Lambda_{k\ell}(b)&\equiv& \int_b^{N}\upd a\,\widetilde f_{k\ell}(b-a)\,{\left[\widetilde A_\ell^{(\Tar)}(a)-\widetilde A_\ell^{(\Sim)}(a)\right]},\qquad
\end{eqnarray}
we have
\begin{equation}\label{eq:Delta2_expression}
\Delta_2(\cdot)=-\!\sum_{k\ell}\!\int_0^{N}\!\!\!\!\!\upd b\,{\left[\widetilde A_k^{(\Tar)}\!(b)\Lambda_{k\ell}(b)
+\!\overline\Lambda_{k\ell}(b)^\dagger\widetilde A_k^{(\Sim)}\!(b)\right]}(\cdot)
\end{equation}
as desired. The expression for $\Delta_1$ can be found in a similar manner. That $\Delta_3(\cdot)=[\Delta_2(\cdot)]^\dagger$ is apparent from Eq.~\eqref{eq:ENImu}.
\end{widetext}

\section{$D_0(c;x)$ for the single-gate exact simulator}\label{app:D0}
We first gather a few basic relations we will use over and over to approximate various terms in our expressions when $|x|,\epmax\ll 1$. In what follows, we assume that $N$ is not large, such that $N|x|$ and $N\epmax$ remain $\ll 1$. Here, $z$ is a variable taken to be small, i.e., $|z|,N|z|\ll 1$.
\begin{eqnarray}\label{eq:ApproxExp}
\upe^{\upi z}-1&=&\upi z+\tfrac{1}{2}(\upi z)^2+O(z^3)\nonumber\\
\frac{\upe^{\upi z}-1}{\upi z}&=&1+\tfrac{1}{2}(\upi z)+O(z^2)\nonumber\\
\frac{1-\upe^{\upi zN}}{1-\upe^{\upi z}}&=&N{\left[1+\tfrac{\upi}{2}(N-1)z+O(z^2)\right]}
\end{eqnarray}

We begin with Eq.~\eqref{eq:D0}, repeated here for the reader's convenience:
\begin{equation}
D_0(c;x)=\frac{\upi{\left[\epsilon{\left(1-\upe^{\upi xc}\right)}-x{\left(1-\upe^{\upi \epsilon}\right)}+\upe^{\upi xc}x{\left(1-\upe^{\upi\epsilon(1-c)}\right)}\right]}}{x(x-\epsilon)}.
\end{equation}
Consider first the situation where $|\bar x|,x_c\ll 1$, such that the spectral function is significant only for $|x|\ll1$. In addition, we have the simulation assumption that $|\epsilon|\ll1$, and note that $c\in(0,1]$. In this limit, to linear order in both $\epsilon$ and $x$, the numerator of $D_0(c;x)$ takes the approximate form
\begin{eqnarray}
&&-\epsilon x c{\left\{{\Bigl[-\frac{\upi xc}{2}+O(x^2)\Bigr]}+{\Bigl[\upi\epsilon{\left(1-\frac{c}{2}\right)}+O(\epsilon^2)\Bigr]}\right.}\\
&&\quad{\left.-\upi x(1-c){\Bigl[1+\frac{\upi\epsilon(1-c)}{2}+O(\epsilon^2)\Bigr]}{\Bigl[1+\frac{\upi xc}{2}+O(x^2)\Bigr]}\right\}}.\nonumber
\end{eqnarray}
If $|x|\ll\epmax$ (regime I), keeping only the leading terms, the numerator becomes $-\epsilon x c{\left[\upi\epsilon(1-\tfrac{c}{2})\right]}$. Together with the denominator, given by $-\epsilon x[1+O(x/\epsilon)]$, we have
\begin{equation}
D_0(c;x)\simeq\upi\epsilon c{\bigl(1-\tfrac{c}{2}\bigr)}\qquad {\left(\textrm{regime I}\right)},
\end{equation}
independent of $x$.
If instead, we have $\epmax\ll |x|$ (regime II), the numerator is $\simeq-\epsilon x c{\left[-\frac{\upi xc}{2}-\upi x(1-c)\right]}=\upi \epsilon x^2 c(1-\tfrac{c}{2})$. Together with the denominator, which is now $x^2[1+O(\epsilon/x)]$, we have again,
\begin{equation}\label{eq:D0regime2}
D_0(c;x)\simeq\upi\epsilon c{\bigl(1-\tfrac{c}{2}\bigr)}\qquad {\left(\textrm{regime II}\right)},
\end{equation}
the same expression as in regime I.
For regime III, where the spectral function is significant only for $|x|\gg 1$, the numerator of $D_0(c;x)$ is $\simeq\upi\epsilon x{\left[1-(1-c)\upe^{\upi xc}+O(\epsilon)\right]}$.
This, with the denominator, which is $\tfrac{\upi}{x^2}\bigl[1+O(\epsilon/x)\bigr]$, we have
\begin{equation}\label{eq:D0regime3}
D_0(c;x)\simeq -\frac{\epsilon}{x}{\left[1-(1-c)\upe^{\upi xc}\right]}\quad~~{\left(\textrm{regime III}\right)}.
\end{equation}
\smallskip

\section{$\Delta_i(\,\cdot\,)$s for the single-gate exact simulator}\label{app:Deltai}
Here, we find expressions for the $\Delta_i(\,\cdot\,)$s for the single-gate exact simulator $\sS_1$, in the limit of $R=\tau_g/T\rightarrow 0$. For $\Delta_1(\cdot)$, we need the sum
\begin{equation}
\Lambda_{k\ell}(b)+\overline{\Lambda}_{k\ell}(b)=\sum_\epsilon \widetilde A_\ell(\epsilon)\sum_{q=0}^{N-1}\upe^{-\upi\epsilon(q+1)}I_{k\ell;q}(b;1),
\end{equation}
with $I_{k\ell;q}(b;1)=\int_{-\infty}^\infty \upd x J_{k\ell}(x)\upe^{-\upi x(b-q)}D_0(1;x)$. Note that $D_0(1;c)$ contains a dependence on $\epsilon$ [see Eq.~\eqref{eq:D0}] that we are not writing explicitly, to not overburden the notation. We repeat here the expressions for the dimensionless interaction-picture $A$ operators, in the limit of $R\rightarrow 0$,
\begin{eqnarray}
\widetilde A^{(\Tar)}_\ell(b)&=&\sum_\epsilon\widetilde A_\ell(\epsilon)\upe^{-\upi\epsilon b}\nonumber\\
\textrm{and}\qquad\widetilde A^{(\Sim)}_\ell(b)&=&\sum_\epsilon\widetilde A_\ell(\epsilon)\upe^{-\upi\epsilon (\lfloor b\rfloor+1)},
\end{eqnarray}
and note that $\widetilde A_\ell(\epsilon)^\dagger =\widetilde A_\ell(-\epsilon)$.
Then, $\Delta_1(\,\cdot\,)$ is given by
\begin{widetext}
\begin{eqnarray}
\Delta_1(\,\cdot\,)&=&\sum_{k\ell}\sum_{\epsilon\epsilon'}\widetilde A_\ell(\epsilon)(\,\cdot\,)\widetilde A_k(\epsilon')\sum_{q=0}^{N-1}\upe^{-\upi\epsilon(q+1)}\int_{0}^N\upd b\int_{-\infty}^\infty\upd x\,\widetilde J_{k\ell}(x)\upe^{-\upi x(b-q)}D_0(1;x)\upe^{-\upi\epsilon'b}\nonumber\\
&&\quad+\sum_{k\ell}\sum_{\epsilon\epsilon'}\widetilde A_k(\epsilon')(\,\cdot\,)\widetilde A_\ell(-\epsilon)\sum_{q=0}^{N-1}\upe^{\upi\epsilon(q+1)}\int_{0}^N\upd b\int_{-\infty}^\infty\upd x\,\widetilde J_{k\ell}(x)^{\!*}\upe^{\upi x(b-q)}D_0(1;x)^{\!*}\upe^{-\upi\epsilon'(\lfloor b\rfloor+1)}.
\end{eqnarray}
The $b$ integrals can be done first (assuming regularity properties of $J_{k\ell}(x)$ for the integration order to not matter):
\begin{equation}
\int_{0}^N\upd b\,\upe^{-\upi xb}\upe^{-\upi\epsilon' b}=\frac{1-\upe^{-\upi (x+\epsilon')N}}{\upi(x+\epsilon')};\quad\quad 
\int_{0}^N\upd b\,\upe^{\upi xb}\upe^{-\upi\epsilon'(\lfloor b\rfloor+1)}=\frac{\upe^{-\upi\epsilon'}(\upe^{\upi x}-1)}{\upi x}\frac{1-\upe^{\upi (x-\epsilon')N}}{1-\upe^{\upi (x-\epsilon')}}\,.
\end{equation}
The sum over $q$ can also be done, with the $q$-sum in the first line of $\Delta_1$ as
\begin{equation}
\sum_{q=0}^{N-1}\upe^{\upi(x-\epsilon) q}=\frac{1-\upe^{\upi(x-\epsilon)N}}{1-\upe^{\upi(x-\epsilon)}},
\end{equation}
the sum of an $N$-term geometric series.
The $q$-sum in the second line of $\Delta_1$ is the complex conjugate of the above sum.
Now, $\Delta_1(\cdot)$ reads as
\begin{eqnarray}\label{eq:Delta1}
\Delta_1(\,\cdot\,)&=&\sum_{k\ell}\sum_{\epsilon\epsilon'}\widetilde A_\ell(\epsilon)(\,\cdot\,)\widetilde A_k(\epsilon')\upe^{-\upi\epsilon}\int_{-\infty}^\infty\upd x\,\widetilde J_{k\ell}(x)D_0(1;x)
\,\frac{1-\upe^{-\upi (x+\epsilon')N}}{\upi(x+\epsilon')}\frac{1-\upe^{\upi(x-\epsilon)N}}{1-\upe^{\upi(x-\epsilon)}}\nonumber\\
&&+\sum_{k\ell}\sum_{\epsilon\epsilon'}\widetilde A_k(\epsilon')(\,\cdot\,)\widetilde A_\ell(-\epsilon)\upe^{\upi\epsilon}\int_{-\infty}^\infty\upd x\,\widetilde J_{k\ell}(x)^{\!*}D_0(1;x)^{\!*} \,\frac{\upe^{-\upi\epsilon'}(\upe^{\upi x}-1)}{\upi x}\frac{1-\upe^{\upi (x-\epsilon')N}}{1-\upe^{\upi (x-\epsilon')}}\frac{1-\upe^{-\upi(x-\epsilon)N}}{1-\upe^{-\upi(x-\epsilon)}}.
\end{eqnarray}
For $\Delta_2(\cdot)$,
putting the expressions for $\Lambda_{kl}$ and $\bar{\Lambda}_{kl}$ into Eq.~\eqref{eq:Delta2_expression}, we have
\begin{eqnarray}\label{eq:Delta2}
&&\Delta_2(\cdot)\nonumber\\
&=&-\sum_{k\ell}\sum_{\epsilon\epsilon^\prime}
\widetilde{A}_k(\epsilon')\widetilde{A}_l(\epsilon)\,(\cdot)\,\upe ^{-\upi\epsilon}\int_{-\infty}^{\infty}\upd x \, \widetilde J_{kl}(x)D_0(1;x)\frac{1-\upe^{-\upi(x+\epsilon')}}{\upi(x+\epsilon')
\left(1-\upe^{\upi(x-\epsilon)}\right)}\left[\frac{1-\upe^{-\upi(x+\epsilon')N}}{1-\upe^{-\upi (x+\epsilon')}}-\frac{1-\upe^{-\upi (\epsilon+\epsilon')N}}{1-\upe^{-\upi(\epsilon+\epsilon')}}\right]
\nonumber\\
&&-\sum_{kl}\sum_{\epsilon\epsilon'}\tilde{A}_l(\epsilon)\tilde{A}_k(\epsilon^\prime)\,(\cdot)\,\upe^{-\upi (\epsilon+\epsilon')}\!\!\int_{-\infty}^{\infty}\!\!\!\upd x\, J_{kl}(x)^* D_0(1;-x)\frac{\upe ^{\upi x}-1}{\upi x \left(1-\upe^{-\upi(x+\epsilon)}\right)}\!{\left[\frac{1-\upe^{-\upi (\epsilon+\epsilon')N}}{1-\upe^{-\upi(\epsilon+\epsilon')}}
-\frac{1-\upe^{\upi(x-\epsilon')N}}{1-\upe^{\upi (x-\epsilon')}}\upe^{-\upi(x+\epsilon)N}\right]}\nonumber\\
&&-\sum_{kl}\sum_{\epsilon\epsilon'}\tilde{A}_k(\epsilon')\tilde{A}_l(\epsilon)\,
(\cdot)\,\upe^{-\upi \epsilon}\frac{1-\upe^{-\upi(\epsilon+\epsilon')N}}{1-\upe^{-\upi(\epsilon+\epsilon')}}
\int_{-\infty}^{\infty}\, \upd x J_{kl}(x)\int_{0}^{1} \upd b' D_0(b^\prime; x)\upe^{-\upi(x+\epsilon')b'} \nonumber\\
&&+\sum_{kl}\sum_{\epsilon\epsilon'}\tilde{A}_l(\epsilon)\tilde{A}_k(\epsilon')\,
(\cdot)\,\upe^{-\upi(\epsilon+\epsilon')}\frac{1-\upe^{-\upi (\epsilon+\epsilon')N}}{1-\upe^{-\upi(\epsilon+\epsilon')}}\int_{-\infty}^{\infty}\, \upd x J_{kl}(x)^* \int_{0}^{1}\, \upd b' D_0(b'; -x)\,\upe^{\upi xb'}.
\end{eqnarray}
\end{widetext}

For $|\epsilon|,|\epsilon'|,|x|\ll 1$, and for $N$ considered as $O(1)$ so that $(x-\epsilon)N,(x\pm\epsilon')N\ll 1$, we can approximate the various exponential terms, to linear order in $\epsilon$, $\epsilon'$ and $x$, using Eq.~\eqref{eq:ApproxExp}.
For regime II, say, where $|\epsilon|, |\epsilon'\ll |x|\ll 1$, we then have
\begin{eqnarray}
\Delta_1(\,\cdot\,)&\simeq&\sum_{k\ell}\sum_{\epsilon\epsilon'}\widetilde A_\ell(\epsilon)(\,\cdot\,)\widetilde A_k(\epsilon')\int_{-\infty}^\infty\upd x\,\widetilde J_{k\ell}(x)\frac{\upi\epsilon N^2}{2}\nonumber\\
&&~+\sum_{k\ell}\sum_{\epsilon\epsilon'}\widetilde A_k(\epsilon')(\,\cdot\,)\widetilde A_\ell(-\epsilon)\int_{-\infty}^\infty\upd x\,\widetilde J_{k\ell}(x)^*\frac{-\upi\epsilon N^2}{2}\nonumber\\
&=&\sum_{k\ell}\sum_{\epsilon\epsilon'}\widetilde A_\ell(\epsilon)(\cdot)\widetilde A_k(\epsilon')\widetilde f_{k\ell}(0)\frac{\upi\epsilon N^2}{2}\nonumber\\
&&\quad+\sum_{k\ell}\sum_{\epsilon\epsilon'}\widetilde A_\ell(\epsilon)(\cdot)\widetilde A_k(\epsilon')\widetilde f_{\ell k}(0)^*\frac{\upi\epsilon' N^2}{2}\nonumber\\
&=&\frac{\upi N^2}{2}\sum_{k\ell}\widetilde f_{k\ell}(0)\sum_{\epsilon\epsilon'}(\epsilon+\epsilon')\widetilde A_\ell(\epsilon)(\cdot)\widetilde A_k(\epsilon'),
\end{eqnarray}
where in the last line, we have used the fact that $\widetilde f_{\ell k}(a)^*=\widetilde f_{k\ell}(-a)$, and replaced $D_0(1;x)$ by the approximate expression of Eq.~\eqref{eq:D0regime2}. A similar analysis for regime I yields the same approximate expression for $\Delta_1(\cdot)$.

For $\Delta_2$ in regimes I and II, again, the various exponential expressions in Eq.~\eqref{eq:Delta2} can be estimated using Eq.~\eqref{eq:ApproxExp}, and one ends up with
\begin{eqnarray}
&&\Delta_2(\,\cdot\,)\nonumber\\
&\simeq& -\sum_{k\ell}\sum_{\epsilon\epsilon'}\widetilde{A}_k(\epsilon')\widetilde{A}_\ell(\epsilon)(\,\cdot\,)\widetilde{f}_{k\ell}(0)\tfrac{\upi }{4}\epsilon N(N-1)\nonumber\\
  &&-\sum_{k\ell}\sum_{\epsilon\epsilon'}\widetilde{A}_\ell(\epsilon)\widetilde{A}_k(\epsilon')(\,\cdot\,)\widetilde{f}_{k\ell}(0)^*\tfrac{\upi}{4}\epsilon N(N+1)\nonumber\\
  &&-\sum_{k\ell}\sum_{\epsilon\epsilon'}  \tilde{A}_k(\epsilon')\tilde{A}_l(\epsilon)\,(\,\cdot\,)\,\widetilde{f}_{kl}(0)\tfrac{\upi}{3}\epsilon N\nonumber\\
  && +\sum_{k\ell}\sum_{\epsilon\epsilon'}\widetilde{A}_\ell(\epsilon)\widetilde{A}_k(\epsilon')\,(\,\cdot\,) \widetilde{f}_{k\ell}(0)^*\tfrac{\upi}{3}\epsilon N\nonumber\\
&=&-\tfrac{\upi}{12} N\sum_{k\ell}\sum_{\epsilon\epsilon'}\widetilde{A}_k(\epsilon')\widetilde{A}_\ell(\epsilon)(\,\cdot\,)\widetilde{f}_{k\ell}(0)\nonumber\\
&&\qquad\qquad\qquad\times{\left[3N(\epsilon+\epsilon')+(\epsilon-\epsilon')\right]}.
\end{eqnarray}
For large $N$, the first term in the brackets above dominates the second one, so that $\Delta_2\sim N^2$. $\Delta_3(\,\cdot\,)=[\Delta_2(\,\cdot\,)]^\dag$ yields the approximate expression for $\Delta_3$.

In regime III, where $|\epsilon|, |\epsilon'|\ll 1\ll |x|, x_c$, the argument in the main text (see the opening paragraph of Sec.~\ref{sec:regime3}) tells us that the $q$-sums in $\Lambda_{k\ell}(b)$ and $\overline\Lambda_{k\ell}(b)$ contain, as significant terms, only those for which $q=\lfloor b \rfloor, \lfloor b \rfloor \pm 1$. One then has, for $p\equiv \lfloor b\rfloor$ such that $b=p+\slashed{b}$,
\begin{eqnarray}
\Lambda_{k\ell}(b)&\simeq&\sum_\epsilon\widetilde A_\ell(\epsilon)\upe^{-\upi\epsilon p}{\left[\Theta(p-1)I_{k\ell;p-1}(b;1)\right.}\nonumber\\
&&\hspace*{3cm}{\left.+\upe^{-\upi\epsilon}I_{k\ell;p}(b;\slashed{b})\right]}\nonumber\\
\overline{\Lambda}_{k\ell}(b)&\simeq&\sum_\epsilon\widetilde A_\ell(\epsilon)\upe^{-\upi\epsilon p}{\left[\Theta(N-2-p)\upe^{-\upi2\epsilon}I_{k\ell;p+1}(b;1)\right.}\nonumber\\
&&\hspace*{0.6cm}{\left.+\upe^{-\upi\epsilon}I_{k\ell;p}(b;1)-\upe^{-\upi\epsilon}I_{k\ell;p}(b;\slashed{b})\right]}
\end{eqnarray}
$\Theta(\cdot)$, as in the main text, is the step function, with the added definition that $\Theta(0)=1$.

$\Delta_1$ in this regime can then be estimated, after some straightforward algebra, to be
\begin{eqnarray}
\Delta_1(\cdot)&\simeq&N\sum_{kl}\sum_{\epsilon\epsilon'}(-\epsilon){\left[\widetilde A_\ell(\epsilon)(\cdot)\widetilde A_k(\epsilon')+\widetilde A_k(\epsilon')(\cdot)\widetilde A_\ell(\epsilon)\right]}\nonumber\\
&&\times\int_{-\infty}^\infty\upd x\widetilde J_{k\ell}(x)\frac{1-\upe^{\upi x}}{\upi x^2}{\left[1+\tfrac{2(N-1)}{N}\cos x\right]}.
\end{eqnarray}
Here, we have set $\upe^{-\upi\epsilon b},\upe^{-\upi\epsilon' b}\simeq 1$, and $D_0(1;x)\simeq-\epsilon/x$ [Eq.~\eqref{eq:D0regime3}].

Likewise, one can estimate $\Delta_2$ in regime III as
\begin{eqnarray}\label{eq:Delta2regime3}
&&\Delta_2(\,\cdot\,)\\
&\simeq&\sum_{k\ell}\sum_{\epsilon\epsilon'}\widetilde A_k(\epsilon')\widetilde A_\ell(\epsilon)(\,\cdot\,)\epsilon\int_{-\infty}^\infty\upd x\widetilde J_{k\ell}(x)\frac{1}{x}\nonumber\\
&&\quad \times{\left\{\tfrac{1}{\upi x}{\left[(1-\upe^{-\upi x})+(N-1)(1-\upe^{-\upi 2x})\right]}-\tfrac{N}{2}\right\}}\nonumber\\
&&+\sum_{k\ell}\sum_{\epsilon\epsilon'}\widetilde A_\ell(-\epsilon)\widetilde A_k(\epsilon')(\,\cdot\,)\epsilon\int_{-\infty}^\infty\upd x\widetilde J_{k\ell}(x)^*\frac{1}{x}\nonumber\\
&&\quad \times{\left\{\tfrac{1}{\upi x}(N-1)\upe^{-\upi x}(1-\upe^{-\upi x})+\tfrac{N}{2}\right\}}\nonumber\\
&\simeq&-\tfrac{1}{2}N\sum_{k\ell}\sum_{\epsilon\epsilon'}\widetilde A_k(\epsilon')\widetilde A_\ell(\epsilon)(\,\cdot\,)\epsilon\int_{-\infty}^\infty\upd x\widetilde J_{k\ell}(x)\frac{1}{x}\nonumber\\
&&-\tfrac{1}{2}N\sum_{k\ell}\sum_{\epsilon\epsilon'}\widetilde A_\ell(\epsilon)\widetilde A_k(\epsilon')(\,\cdot\,)\epsilon\int_{-\infty}^\infty\upd x\widetilde J_{k\ell}(x)^*\frac{1}{x}.\nonumber
\end{eqnarray}
In the last (approximate) equality, we have dropped the $1/x$ terms within the curly braces, since they are small in regime III ($|x|\gg 1$ here), compared to the order-1 $N/2$ terms. Now, the two integrals can be rewritten as,
\begin{eqnarray}
\int_{-\infty}^{\infty}\upd x \frac{\widetilde J_{k\ell}(x)}{x}&=&\,-\upi\int_{-\infty}^{0}\upd a\widetilde{f}_{k\ell}(a)+\upi\pi \widetilde J_{k\ell}(0),\\
\int_{-\infty}^{\infty}\upd x\frac{\widetilde J_{k\ell}(x)^*}{x}&=&\,-\upi\int_{-\infty}^{0}\upd a\widetilde{f}_{k\ell}(-a)^*+\upi\pi \widetilde J_{k\ell}(0)^*.\nonumber
\end{eqnarray}
In regime III, $\widetilde J_{k\ell}(0)$ can be taken to be zero---$\widetilde J$ is significant in regime III only for large $|x|$ values. Hence, we finally have,
\begin{eqnarray}
&&\Delta_2(\,\cdot\,)\\
&\simeq&\tfrac{\upi}{2}N\sum_{k\ell}\sum_{\epsilon\epsilon'}\widetilde A_k(\epsilon')\widetilde A_\ell(\epsilon)(\,\cdot\,)\epsilon\int_{-\infty}^0\upd a\widetilde f_{k\ell}(a)\nonumber\\
&&+\tfrac{\upi}{2}N\sum_{k\ell}\sum_{\epsilon\epsilon'}\widetilde A_\ell(\epsilon)\widetilde A_k(\epsilon')(\,\cdot\,)\epsilon\int_{-\infty}^0\upd a\widetilde f_{k\ell}(-a)^*\nonumber\\
&=&\tfrac{\upi}{2}N\sum_{k\ell}\sum_{\epsilon\epsilon'}\widetilde A_k(\epsilon')\widetilde A_\ell(\epsilon)(\,\cdot\,)(\epsilon+\epsilon')\int_{-\infty}^0\upd a\widetilde f_{k\ell}(a)\nonumber
\end{eqnarray}
Lastly, as usual, $\Delta_3(\,\cdot\,)=\Delta_2(\,\cdot\,)^\dagger$ gives us the approximate expression for $\Delta_3$.

Note that $\Delta_1$ involves terms of order $1/x^2$ in the integrand, which are of the same order as those we dropped in computing $\Delta_2$ [see comment right after Eq.~\eqref{eq:Delta2regime3}]. Hence, in regime III, $\Delta_1$ can be considered negligible compared to $\Delta_2$ and $\Delta_3$.

\bigskip
\section{Glossary}
\label{app:Glossary}

We gather here a list of symbols and notation that will be helpful for the reader to navigate the main text. Throughout the text, we choose units such that $\hbar=1$ and $k=1$ (the Boltzmann's constant).
\begin{itemize}
\setlength{\itemsep}{0pt}
\setlength{\parsep}{0pt}
\setlength{\parskip}{5pt}
\item $T$ is the stroboscopic simulation cycle time, used as the basic unit of time and inverse frequency (or energy) in our analysis.
\item $\omega$ is a transition frequency of the target system;\\
$\ommax\equiv \max|\omega|$ is the largest transition frequency.
\item$\frac{1}{\omega}$ gives a timescale of the target system;\\
$\frac{1}{\ommax}$ gives the smallest timescale of the target.
\item $\nu_c$ is the cutoff frequency of the bath spectral function.
\item $\tauB=\frac{1}{\nu_c}$ is the bath correlation timescale.
\item $\beta$ is the inverse temperature;\\
$\widetilde\beta=\frac{\beta}{T}$ is its dimensionless version.
\item $\eta$ is the system-bath coupling constant for the oscillator bath, appearing in the spectral density;\\
$\widetilde\eta=\eta T^{1-w}$ is its dimensionless version, with $w$ the frequency power in the spectral density.
\item $\epsilon=\omega T=\frac{\textrm{stroboscopic simulation timescale}}{\textrm{timescale of the target}}$
\item $x_c= \nu_cT=\frac{T}{\tauB}=\frac{\textrm{stroboscopic simulation timescale}}{\textrm{timescale of the bath}}$
\item $a_B\equiv 1/x_c=\tau_B/T$.
\item $\frac{\textrm{timescale of the bath}}{\textrm{timescale of the target}}=\frac{\tauB}{1/\omega}=\omega\tauB=\frac{\omega}{\nu_c}=\frac{\epsilon}{x_c}$
\item $\frac{\textrm{thermal energy}}{\textrm{energy scale for the target system}}=\frac{1/\beta}{\omega}=\frac{1}{\beta\omega}=\frac{1}{\widetilde\beta \epsilon}$
\item $\tau_g$ is the time taken to complete the $M$-gate sequence for the DQS.
\item $R_M=\tau_g/T$ is the ratio of the $M$-gate sequence time $\tau_g$ to the simulation cycle time $T$. When the value of $M$ is clear, we often drop the subscript $M$ and simply write $R$.
\item $\sS_M$ is a DQS that uses an $M$-gate sequence for the digital simulation of the target Hamiltonian.
\end{itemize}


%

\end{document}